\documentclass[preprint,aps,nofootinbib,preprintnumbers]{revtex4-1}
\pdfoutput=1

\usepackage[normalem]{ulem}
\usepackage{amsmath,amssymb,amsfonts,dcolumn,color,graphicx,graphics,latexsym,placeins,epsfig}
\usepackage{epsfig}
\usepackage{bm}
\usepackage{slashed}
\usepackage{latexsym}
\usepackage{natbib}
\usepackage{url}
\usepackage{dcolumn}
\usepackage{color}
\usepackage{amsfonts,amssymb,amsmath}
\usepackage{graphicx,epsfig}
\usepackage{psfrag}
\usepackage{subfigure}
\usepackage{tabularx}
\usepackage{hyperref}
\hypersetup{colorlinks=true}
\usepackage{comment}

\newcommand{\be}{\begin{equation}}
\newcommand{\ee}{\end{equation}}
\newcommand{\ba}{\begin{eqnarray}}
\newcommand{\ea}{\end{eqnarray}}

\begin{document}

\title{Scalar and vector dark matter admixed neutron stars with linear and quadratic couplings}
\author{Francesco Grippa$^{1,2,}$\footnote{fgrippa@unisa.it}}
\author{Gaetano Lambiase$^{1,2,}$\footnote{lambiase@sa.infn.it}}
\author{Tanmay Kumar Poddar$^{2,3,}$\footnote{tanmay.k.poddar@durham.ac.uk}}
\affiliation{${}^{1}$ Dipartimento di Fisica E.R. Caianiello, Universit\`a di Salerno, Via Giovanni Paolo II 132 I-84084, Fisciano (SA), Italy}
\affiliation{${}^{2}$ INFN, Gruppo collegato di Salerno, Via Giovanni Paolo II 132 I-84084, Fisciano (SA), Italy.}
\affiliation{${}^{3}$ Institute for Particle Physics Phenomenology (IPPP), Department of Physics, Durham University, Durham DH1 3LE, United Kingdom}

\begin{abstract}
We investigate the effects of dark scalar- and vector-mediated interactions on dark matter admixed neutron stars, employing the two-fluid formalism. We adopt three different nuclear equations of state -- BSk22, MPA1 and APR4 -- to describe the baryonic sector, while the dark component consists of fermionic particles within a relativistic mean field framework. We consider both linear and quadratic scalar interactions with the dark fermion, including a quartic self-interaction in the latter case. The parameters of the dark matter models are inferred via a Bayesian analysis that incorporates data from NICER observations and binary neutron star merger detections. The neutron star configurations obtained from the selected model parameters develop dark matter cores, leading to more compact objects with smaller masses and radii. Our findings suggest that scalar interactions generally have a weaker impact on the stellar structure compared to vector-mediated ones, though quantitative differences arise. In particular, quadratic scalar couplings suppress the net attractive interaction, allowing for larger dark matter fractions to be accreted. We also compute the sound speed of DM, finding that the scalar and quadratic interactions modify the stiffness of the dark equation of state while respecting causality: vector repulsion enhances the sound speed, whereas scalar attraction tends to soften it. We compare our results with GW1708017, GW190425 and NICER data and constrain DM couplings and mass.
\end{abstract}

\maketitle
\section{Introduction}

Dark matter (DM) constitutes a fundamental component of our universe, with an energy density roughly five times greater than that of visible matter \cite{Planck:2018vyg}. Its presence has been confirmed from the galactic scale to the cosmological scale \cite{Catena:2009mf,Sofue:2000jx,Markevitch:2003at,Liddle:1993fq}. DM is primarily cold and non-relativistic, playing a crucial role in the formation of cosmic structures. It is established that DM interacts only through gravitational forces. However, if DM interacts very weakly with Standard Model (SM) particles allowed by the current data, such interactions can be tested by various ongoing and upcoming laboratories as well as astrophysical and cosmological experiments. The mass of DM spans a wide range, from as low as $10^{-22}~\mathrm{eV}$ (fuzzy DM) \cite{Hu:2000ke, Hui:2016ltb, Ferreira:2020fam} to about a fraction of the mass of the Sun (primordial black hole) \cite{Carr:2016drx, Carr:2020gox}. Given the extremely feeble interactions of DM with SM particles, aside from gravity, compact and highly dense astrophysical objects such as neutron stars (NSs) and black holes (BHs) provide exceptional cosmic laboratories for probing DM properties. These environments offer sensitivity complementary to that of terrestrial direct-detection experiments \cite{Raffelt:1996wa,Bramante:2023djs,Grippa:2024ach}.

The internal structure of NSs remains largely unknown, and several Equations of States (EoSs) have been developed to explore the stellar properties of NSs \cite{Lattimer:2021emm}. These EoSs are constructed from the characteristics of interactions among the fundamental constituents, shaping the internal structure of the NS by determining its pressure and energy density. Their gradients are described by the Tolman-Oppenheimer-Volkoff (TOV) equations \cite{Oppenheimer:1939ne, Rutherford:2022xeb}. The stress-energy tensor of the baryonic fluid inside a NS varies with different EoSs. The presence of DM within the NSs can alter the stellar structure as well. This can be modeled by formulating a similar set of TOV equations for the DM fluid.

Gravitational wave (GW) detections, including events such as GW170817 \cite{LIGOScientific:2017vwq}, GW190814 \cite{LIGOScientific:2020zkf}, and GW190425 \cite{LIGOScientific:2020aai} from the LIGO and Virgo collaborations, provide insights into the measurements of NS masses and radii. X-ray observations from NICER \cite{Miller:2019cac,Miller:2021qha,Riley:2019yda,Riley:2021pdl,Raaijmakers:2019dks}, along with optical \cite{Romani:2021xmb} and radio \cite{Antoniadis:2013pzd,Fonseca:2021wxt} measurements of rotating pulsars further constrain the internal properties of these stars. The presence of DM within the NSs can influence their mass-radius (M-R) relationship, and parameters related to DM can be constrained using these observational data.

In addition, the tidal deformability $\Lambda$ can be affected by the presence of DM. In binary systems, $\Lambda$ measures how much a star deforms due to the gravitational field of its companion \cite{Binnington:2009bb}. Lower tidal deformability indicates that the NS is more compact and experiences less deformation. The tidal field induces a quadrupole moment, quantifying the distortion from spherical symmetry. When DM is present as a halo or a core, it acts like an additional tidal field, influencing the measurements of $\Lambda$.

Another important quantity that characterizes a NS is the sound speed $c_s$, namely the velocity at which linear perturbations propagate in a uniform fluid \cite{Rezzolla:2013dea}. The sound speed $c_s$ can be computed as the derivative of the pressure with respect to the energy density. Near the core, the pressure increases sharply with density, resulting in a stiff EoS and a high sound speed. On the contrary, at the star's surface, the sound speed decreases due to lower density. To satisfy the causality condition, the sound speed must remain below the speed of light. The presence of DM within the NS alters the pressure and density, thereby affecting $c_s$.

The typical mass of a NS ranges from $1.4$ to $2.5~M_\odot$ and its radius can vary between $10$ and $15~\mathrm{km}$ \cite{Chandrasekhar:1939, Nattila:2022evn}, whereas its tidal deformability generally falls within $[100-1000]$. However, these values are highly dependent on the choice of the EoS. The binary neutron star (BNS) merger event GW170817 places a limit on the tidal deformability parameter for an average $1.4~M_\odot$ NS at $\Lambda=190^{+390}_{-120}$ \cite{LIGOScientific:2018cki}.  Another detected coalescence, GW190425, sets an upper bound on the effective deformability parameter for the binary system at $\Lambda \leq 600$ and $\Lambda\leq 1100$, depending on the initial spin prior and masses \cite{LIGOScientific:2020aai}. Since the imprint of $\Lambda$ is encoded in the GW emission \cite{Zhao:2018nyf}, the measurement of the tidal deformability can be useful to infer the properties of the NS \cite{Raithel:2018ncd, Bose:2017jvk} and to put constraints on the EoS \cite{Baiotti:2019sew}. The maximum TOV mass of a NS is crucial in validating or ruling out some EoSs. Several multi-messenger observations predict the maximum TOV mass is around $2.49-2.52~M_\odot$ \cite{Ai:2023ykc}. NICER data provide the radius of PSR J0740-6620 as $12.39^{+1.30}_{-0.98}~\mathrm{km}$ \cite{Riley:2021pdl} and PSR J0030+0451 as $12.71^{+1.14}_{-1.19}~\mathrm{km}$ \cite{Riley:2019yda}.

Depending on its mass, relative abundance, and self-interaction, DM within a NS can either accumulate to form a core or behave as a halo around the star. Typically, a DM core leads to a softer EoS, whereas the formation of a surrounding DM halo stiffens the EoS.

The DM within a NS can be symmetric \cite{Kouvaris:2007ay,Perez-Garcia:2011tqq}, asymmetric \cite{Shelton:2010ta,Petraki:2013wwa,Kouvaris:2015rea}, and can be either bosonic \cite{Giangrandi:2022wht,RafieiKarkevandi:2021hcc,Karkevandi:2021ygv} or fermionic \cite{Narain:2006kx,Goldman:2013qla}. In all DM models, the challenge of characterizing a  dark matter admixed neutron star (DANS) is selecting the proper parameters, particularly the couplings and the masses. Unlike hadronic matter models, which can be tightly constrained by nuclear and astrophysical observations \cite{Lattimer:2006xb, Tsang:2023vhh}, no similar limits exist for DM. In principle, the parameters of the dark sector can vary widely while still providing acceptable M-R relations of DANSs \cite{Xiang:2013xwa}. However, this flexibility can lead to a large number of possible configurations, reducing the predictive power of the model. To mitigate this ambiguity, Bayesian analysis is often employed, as it helps identifying optimized parameters by incorporating the latest experimental and empirical data. The Bayesian framework has been extensively applied to estimate the parameters of nuclear EoSs \cite{Jiang:2022tps, Zhu:2022ibs, Huang:2023grj, Huang:2024rvj}, and it has recently been adopted in the context of DANSs \cite{Das:2020ecp, Koehn:2024gal}.

In this paper, we study DANSs where the baryonic matter (BM) is coupled to DM only by gravity. To do that, one has to model both the BM and the DM with a proper EoS. Many EoSs have been proposed for the baryonic side (see~\cite{Burgio:2021vgk}, for a review). Due to the strict constraints mentioned above, these models share a similar qualitative behavior. As a consequence, different BM EoSs provide slightly different properties of DANSs when in combination with the same DM model. On the contrary, the nature of DM and its EoS are entirely unknown. Several studies have been performed: DM is assumed as a free Fermi gas in \cite{Leung:2011zz, Ivanytskyi:2019wxd};  bosonic and fermionic DM have been studied in \cite{Ellis:2018bkr, Shakeri:2022dwg, Karkevandi:2024vov} and \cite{Liu:2023ecz, Liu:2024rix}, respectively. Ultralight \cite{Diedrichs:2023trk}, mirror \cite{Kain:2021hpk} and fuzzy DM \cite{Rezaei:2023iif} have been also considered. An alternative approach has been proposed in \cite{Guha:2021njn, Guha:2024pnn, Sen:2021wev}, where a feeble non-gravitational interaction between the two fluids in DANSs is considered.

We extend our analysis of DANSs with fermionic DM, treated within a relativistic mean field (RMF) model. In particular, we describe the BM via three different EoSs (BSk22 \cite{Goriely:2013xba, Pearson:2018tkr}, MPA1 \cite{Muther:1987xaa}, APR4 \cite{Akmal:1998cf}) to assess the impact of the nuclear sector, whereas in the dark sector, we consider that the DM is composed of fermions interacting with each other by dark mediators. We chose a vector mediator as well as a scalar mediator, where the scalar can have either linear or quadratic interaction with the dark fermions. The linear scalar interaction has sometimes been considered, although often in combination with a similar Lagrangian approach for both BM and DM \cite{Das:2020ecp, Routaray:2023spb}. The quadratic scalar interaction case is instead a new scenario that we consider here to further explore the influence of the interaction among DM fermions mediated by dark scalars. We exploit a Bayesian parameter optimization method to choose the parameters of both models. By solving the corresponding 2-fluid TOV equations, we aim to investigate the effects of such interactions on the stellar properties, such as the M-R and the $\Lambda$-M relations. Our findings are then compared with experimental constraints from isolated or binary NSs. 

The paper is structured as follows. In Section~\ref{sec:numerical_setup}, we describe the numerical setup of the general relativistic framework used to solve the stellar structure of DANSs. We also outline the formalism for calculating the tidal deformability of a general compact star. In Section~\ref{sec:EoS}, we describe the EoSs for both BM and DM. Within the Lagrangian approach, we introduce linear and quadratic couplings of dark scalars with DM fermions. We follow the RMF prescription to calculate observables such as the sound speed, while the DM parameters are estimated via Bayesian inference. In Section~\ref{sec:mrtd}, we display the resulting mass-radius (M-R) and tidal deformability-mass ($\Lambda$-M) relations, that we compare with the latest constraints posed by GW170817, GW190814, as well as NICER observations of PSR J0740+6620 and PSR J0030+0451. We conclude and summarize our findings in Section~\ref{sec:conclusions}.

We use $c$ (speed of light in vacuum) $=\hbar$ (reduced Planck's constant) $=G$ (Newton's Gravitation constant) $=1$ and choose signature of the metric $\mathrm{diag}(-1,+1,+1,+1)$ unless stated otherwise.

\section{Numerical setup of dark matter admixed neutron star}
\label{sec:numerical_setup}

The general relativistic metric of a static spherically symmetric spacetime is given as \cite{Tolman:1939jz}
\begin{equation}
\label{eq:metric_spherically_simmetric}
    ds^2 = -e^{\nu(r)}dt^2 + \frac{dr^2}{1-2m(r)/r} + r^2(d\theta^2 + sin^2\theta d\phi^2)\, ,
\end{equation}
where $m(r)$ is the enclosed mass within a stellar volume of radius $r$ and $\nu(r)$ is the metric function which decouples in the static case. To study the properties of DANSs, we assume both BM and DM are perfect fluids, and they couple with each other through gravitational interaction only.
The stress-energy tensors of BM and DM are given by \cite{Misner:1973prb}

\begin{equation}
    \label{eq:stress_energy_tensor}
    T_i^{\mu\nu} = (\rho_i + P_i) u^\mu u^\nu + P_i \eta^{\mu\nu} \, , \\
\end{equation}
where $\rho_i$ and $P_i$ ($i$ = BM, DM) are the energy densities and the pressures of the two fluids, respectively. The total stress-energy tensor of the DANS can be written as $T_\mathrm{tot}^{\mu\nu} = T_\mathrm{BM}^{\mu\nu} + T_\mathrm{DM}^{\mu\nu}$, where the conservation law separately holds for each fluid, i.e., $\nabla_{\mu} T_{i}^{\mu\nu} = 0$. The coupled TOV equations, describing the hydrostatic equilibrium of a single, non-rotating NS composed of two gravitationally interacting fluids, are obtained by solving the Einstein field equations $G_{\mu\nu} = 8 \pi T_{\mu\nu}$ for the metric given in Eq.~\eqref{eq:metric_spherically_simmetric} as \cite{Oppenheimer:1939ne}

\begin{align}
\label{eq:TOV}
    &\frac{dP_i}{dr} = - (P_i + \rho_i) \frac{4 \pi r^3 P_\mathrm{tot} + m(r)}{r(r-2m(r))} \, , \\
\label{eq:enclosed_mass}
    &\frac{dm(r)}{dr} = 4 \pi \rho_\mathrm{tot} r^2 \, ,
\end{align}

where $P_\mathrm{tot} = P_\mathrm{BM} + P_\mathrm{DM}$, $\rho_\mathrm{tot} = \rho_\mathrm{BM} + \rho_\mathrm{DM}$ are the total pressure and energy density respectively. We also need an EoS $P_i=P_i(\rho_i)$ (in the barotropic approximation \cite{Bauswein:2012ya}) to formulate a deterministic system. The three equations (Eqs.~\eqref{eq:TOV},~\eqref{eq:enclosed_mass} and the EoS) have to be solved numerically by first fixing the central density $\rho_c$ of the star at $\sim [10^{14} - 10^{15}] \; \mathrm{g/cm^3}$. Equations.~\eqref{eq:TOV}--\eqref{eq:enclosed_mass} are then integrated from the center of the star up to the surface, defined as the locus of points where $P_\mathrm{tot}(r=R)=0$. Note that, solving the equations for two-fluid systems requires fixing\footnote{Fixing $f_\mathrm{DM}$ means to assume a specific DM accumulation mechanism that can be highly model-dependent. A more consistent approach treats $f_\mathrm{DM}$ as a variable, yielding a two-dimensional solution space rather than a single curve \cite{Hippert:2022snq}. Despite being more flexible, this complicates direct comparisons with pure NS models.} an additional parameter, namely the DM fraction $f_\mathrm{DM}$ within the DANS. While a few works \cite{Das:2020ecp} express it in terms of the central energy densities of the two components, we adopt the most common definition \cite{Panotopoulos:2017idn, Ivanytskyi:2019wxd, Karkevandi:2021ygv, Giangrandi:2022wht}, establishing that $f_\mathrm{DM}$ is the ratio between the DM mass and the total mass of the star given as

\begin{equation}
    \label{eq:DM_fraction_definition}
    f_\mathrm{DM} = \frac{M_\mathrm{DM} }{M} = \frac{M_\mathrm{DM} (R_\mathrm{DM})}{M_\mathrm{BM} (R_\mathrm{BM}) +M_\mathrm{DM} (R_\mathrm{DM})} \, ,
\end{equation}

where, in the last equality, we explicitly specify that the DM and the BM masses are enclosed within the radii $R_\mathrm{DM}$ and $R_\mathrm{BM}$, respectively.

For our analysis, we adapt the public \verb|Python| code \cite{Collier:2022cpr} to study the properties of DANS. We consider three nuclear EoSs (see Section~\ref{sec:EoS_baryonic_matter}) for the baryonic sector, namely BSk22 (as in \cite{Collier:2022cpr}, where DM is however described via a different model), MPA1 and APR4. The DM EoSs are derived from a Lagrangian approach that includes dark vector- and scalar-mediated interactions--either linear or quadratic between DM fermions. We also define the dimensional tidal deformability parameter of a single NS as \cite{Hinderer:2007mb}

\begin{equation}
\label{eq:Lambda}
    \Lambda = \frac{2}{3} k_2 \left ( \frac{R}{ M}\right)^5 \, ,
\end{equation}
where $k_2$ is the $\ell = 2$ tidal Love number \cite{Flanagan:2007ix, Damour:2009vw} with a typical range from $0.05$ to $0.15$ \cite{Hinderer:2009ca}. The Love number is computed as \cite{Hinderer:2007mb}

\begin{eqnarray}
    k_2 &=& \frac{8C^5}{5} (1-2C)^2 [2 + 2C (y_R - 1) - y_R]
    \Big\{2C(6-3y_R + 3C(5y_R-8))+
    4C^3\big[13-11y_R + \nonumber \\
    & & C(3y_R-2)+2C^2(1+y_R)\big] 
    +3(1-2C)^2 \big[2-y_R+2C(y_R-1)\big]\mathrm{\ln}(1-2C) \Big\}^{-1} \,,
\label{eq:k_2}
\end{eqnarray}
where $C\equiv M/R$ is the compactness parameter and $y_R \equiv y (r=R)$ is the solution of the two-fluid differential equation \cite{Postnikov:2010yn, Perot:2020gux}

\begin{equation}
\label{eq:y_R}
    r \frac{dy(r)}{dr} + y(r)^2 + y(r)F(r) + r^2Q(r) = 0\, ,
\end{equation}
where
\begin{align}
\label{eq:F(r)}
    F(r) &= \frac{r-4 \pi r^3(\rho_\mathrm{tot} - P_\mathrm{tot})}{r-2m(r)} \, , \\
\label{eq:Q(r)}
    Q(r) &= \frac{4 \pi r [5 \rho_\mathrm{tot} + 9P_\mathrm{tot} + \sum_{i} \frac{\rho_i + P_i}{\partial P_i / \partial \rho_i}  - \frac{6}{4 \pi r^2}}{r - 2m(r)} - 4 \left [ \frac{m(r) + 4 \pi r^3 P_\mathrm{tot}}{r^2(\-2m(r)/r)}\right]^2 \, .
\end{align}
Equations.~\eqref{eq:F(r)}--\eqref{eq:Q(r)} account for the presence of the DM fluid within the stellar volume, whereas the evolution of \textbf{$y(r)$} in Eq.~\eqref{eq:y_R} remains unchanged outside it.

\section{Equations of state for a dark matter admixed neutron star}
\label{sec:EoS}

A NS is fully characterized by its EoS, whose exact form remains, however, uncertain. Starting from several different formulations \cite{Burgio:2021bzy}, many models have been proposed to describe the behavior of matter at super-nuclear densities $n > n_{\mathrm{nuc}} = 0.16 \; \mathrm{fm^{-3}}$. In the following, we describe the EoSs that we employ for the two coexisting fluids.

\subsection{Baryonic matter}
\label{sec:EoS_baryonic_matter}

\begin{figure*}[h]
\centering
\includegraphics[width=\textwidth]{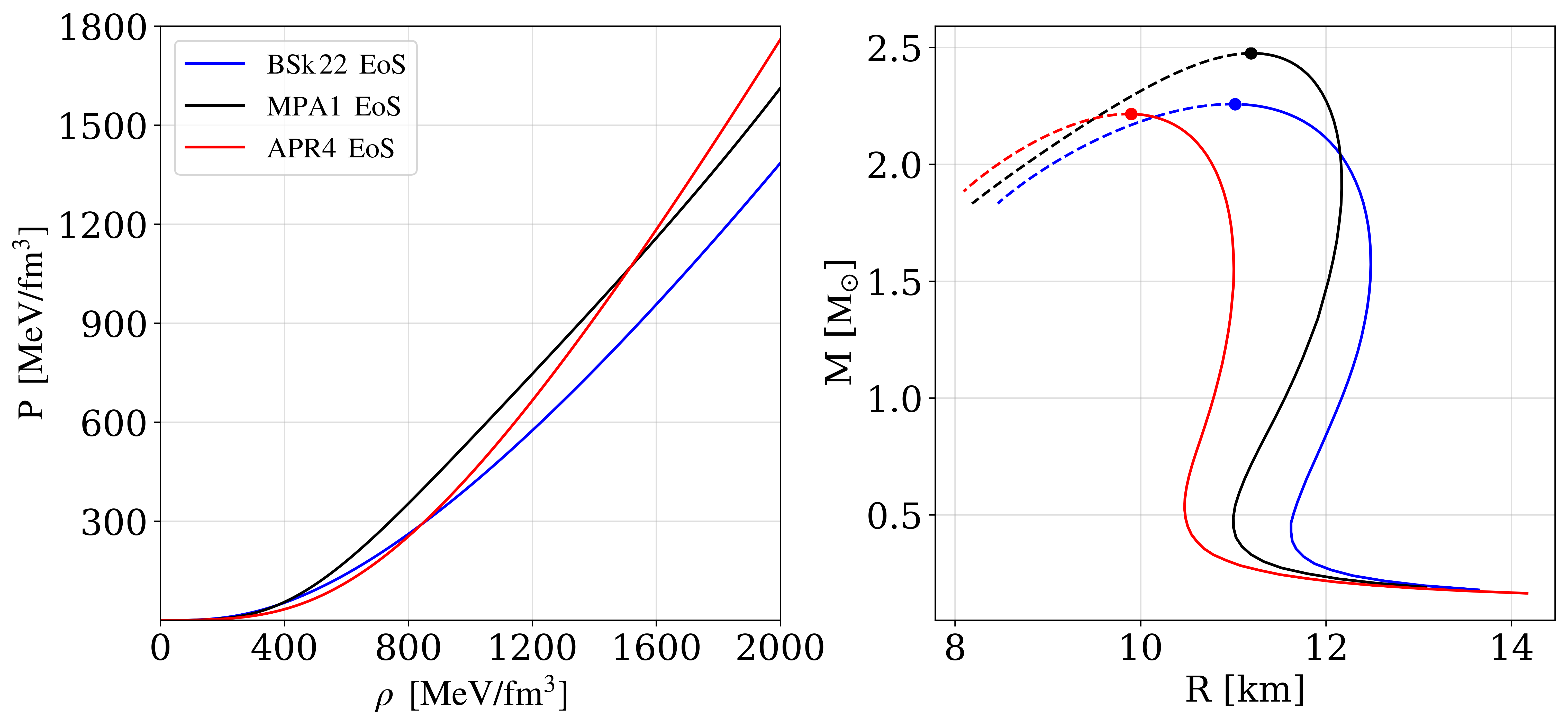}
   \caption{\label{fig:nuclear_EoSs} Pressure-energy density (left panel) and mass-radius (right panel) relations for the nuclear EoSs that we adopt to describe the baryonic sector. BSk22 is the softest, with a lower maximum mass than MPA1 and a greater $R_\mathrm{TOV}$ than APR4. MPA1 is stiffer than APR4 at densities typically reach inside a NS ($\rho \leq 1400 \; \mathrm{MeV/fm^3} \simeq 2.5 \times 10^{15} \; \mathrm{g/cm^3}$).}
\end{figure*}

As mentioned in Section~\ref{sec:numerical_setup}, we describe the BM via three different nuclear EoSs. In particular, we choose the Brussels-Montreal energy density functional BSk22 \cite{Goriely:2013xba, Pearson:2018tkr, Collier:2022cpr}. BSk22 EoS is based on non-relativistic Skyrme interaction, and its parameters are determined from the 2016 Atomic Mass Evaluation \cite{Wang:2021xhn}. We also select different EoSs such as MPA1 and APR4. The former is derived from the relativistic Brueckner-Hartree-Fock approach \cite{Muther:1987xaa}; the latter originates from the variational method proposed in \cite{Akmal:1998cf}.

The EoSs are reported in FIG.~\ref{fig:nuclear_EoSs}, where we display the pressure-energy density relation (left panel) and the resulting mass-radius curves (right panel) computed through Eqs.~\eqref{eq:TOV}-\eqref{eq:enclosed_mass}. Dashed lines represent the unstable stars branch. The greater stiffness of MPA1 (black line) yields a larger maximum mass $M_\mathrm{TOV}^\mathrm{M} = 2.48 M_\odot$ which corresponds to $R_\mathrm{TOV}^\mathrm{M} = 11.2$ km. BSk22 (blue line) and APR4 (red line) have similar maximum masses ($M_\mathrm{TOV}^\mathrm{B} = 2.26 M_\odot$ and $M_\mathrm{TOV}^\mathrm{A} = 2.21 M_\odot$, respectively) but differ in the associated radii ($R_\mathrm{TOV}^\mathrm{B} = 11.02$ km and $R_\mathrm{TOV}^\mathrm{A} = 9.9$ km).

To gain faster convergence, we exploit the analytical fits of \cite{Gungor:2011vq, Potekhin:2013qqa} to implement these EoSs in our code. This treatment slightly smooths the EoSs. However, the effect is small and negligible. We indeed estimate the differences in the maximum masses between the original formulation and the analytical fit to be only $\lesssim 0.03 \,  M_{\odot}$.

\subsection{Dark matter}
\label{sec:EoS_dark_matter}

For the dark sector, we assume the existence of self-interacting dark fermions similar to nucleons which interact with each other by exchanging dark mediators, namely a scalar $\phi(x)$ and a vector $V_\mu(x)$. 

In what follows, we derive the mean values of the fields from the given Lagrangians. We consider scenarios where the scalar field mediator couples linearly and quadratically with the dark fermion.

\subsubsection{Scalar linear coupling}
\label{sec:linear_scalar_coupling}

In the \emph{linear coupling} scenario, we consider a universal model to construct the DM Lagrangian according to the covariance principle. In particular, a neutral dark scalar meson $\phi$ couples to the DM fermion $\overline{\Psi}$ through $g_{s_1}\phi\overline{\Psi}\Psi$ and a neutral dark vector meson $V_\mu$ couples to the conserved DM current through $g_v \overline{\Psi} \gamma_\mu \Psi  V^\mu$. Here, $g_{s_1}$ and $g_v$ denote the scalar and vector coupling strengths with the DM candidate, respectively. The complete Lagrangian for the DM sector is given by
\begin{eqnarray}
    \mathcal{L} &= &\overline{\Psi} (x) \left[ \gamma_{\mu} \left ( i \partial^{\mu}- g_v V^{\mu}(x) \right) - \left(M_\mathrm{D} - g_{s_1} \phi(x) \right) \right] \Psi (x) + \nonumber \\ &+& \frac{1}{2} \big( \partial_{\mu} \phi (x) \partial^{\mu} \phi (x) - m_s^2 \phi^2(x) \big) - \frac{1}{4} V_{\mu\nu}V^{\mu\nu} + \frac{1}{2} m_v^2 V_\mu (x) V^\mu (x),
    \label{eq:linear_lagrangian}
\end{eqnarray}
where $V_{\mu\nu} = \partial_{\mu}V_\nu(x) - \partial_{\nu}V_\mu(x)$, $M_\mathrm{D}$ denotes the mass of the dark fermion candidate, and $m_s$ and $m_v$ denote the masses of the dark scalar and vector mesons, respectively. We adopt the parametrization in \cite{Banerjee:2022sqg} and redefine the vector and the scalar couplings as
\begin{align}
\label{eq:coupling_vector}
    g_v = \frac{d_v  M_\mathrm{D}}{\widetilde{M}_p}, \hspace{0.5cm} g_{s_1} = \frac{d_{s_1}  M_\mathrm{D}}{\widetilde{M}_p} \, ,
\end{align}

where $\widetilde{M}_p = 2.4 \times 10^{18}$ GeV is the reduced Planck mass, and both $d_v$ and $d_{s_1}$ are dimensionless. In what follows, we consider $d_v$ and $d_{s_1}$ as free parameters of our models rather than $g_v/m_v$ and $g_{s_1}/m_s$, which have been studied in \cite{Das:2020ecp}.

The exchange of mesons is such that an effective Yukawa potential arises as \cite{Mukhopadhyay:2016dsg}
\begin{equation}
\label{eq:yukawa_potential}
    V(r) = \frac{g_{v}^2}{4\pi} \frac{e^{-m_v r}}{r} - \frac{g_{s_1}^2}{4\pi} \frac{e^{-m_s r}}{r} \, .
\end{equation}
The behavior of the potential at small and large distances depends on the couplings $g_i$ and the masses $m_i$ of the dark mediators. In principle, there is no unique way to choose these values. This happens because the DM Lagrangian yields an EoS which only depends on the ratios $c_i = g_i / m_i$.

In Section~\ref{sec:bayesian_analysis}, we discuss the selection of the parameters of the DM EoSs, assuming a priori knowledge, exploiting a Bayesian framework. For now, we consider the case $m_s << m_v$. This condition is intended to produce a potential that is attractive at large distances and repulsive at short distances \cite{Xiang:2013xwa}. 

We outline the crucial steps in deriving the EoS within the RMF framework \cite{Shen:1998gq}. The Lagrangian in Eq.~\eqref{eq:linear_lagrangian} leads to the following equations of motion

\begin{align}
\label{eq:eq_motion_psi}
   &\left [ \gamma_\mu \big( i\partial^\mu - g_v V^\mu (x) \big) - \big( M_\mathrm{D} - g_{s_1} \phi(x)\big) \right] \Psi (x) = 0 \\
\label{eq:eq_motion_phi_linear}
   &\partial_\mu \partial^\mu \phi(x) + m_s^2 \phi(x)  =  g_{s_1} \overline{\Psi} (x) \Psi (x) \,, \\
\label{eq:eq_motion_v}
    &\partial_\nu V^{\mu\nu} (x) + m_v^2 V^\mu (x) = g_v \overline{\Psi} (x) \gamma^\mu \Psi (x)  \,.
\end{align}

In the RMF prescription, the mediator field operators are replaced by their ground state expectation values as $\left \langle \phi(x) \right \rangle = \phi_0$, $\left \langle V_\mu(x) \right \rangle = V_0$, which can be computed using Eqs.~\eqref{eq:eq_motion_psi}, \eqref{eq:eq_motion_phi_linear}, and \eqref{eq:eq_motion_v} as \cite{Diener:2008bj}
\begin{align}
\label{eq:exp_value_phi_linear}
     \phi_0 &= \frac{g_{s_1}}{m_s^2} \langle \overline{\Psi} \Psi \rangle  = \frac{g_{s_1}}{m_s^2} \frac{{M_\mathrm{D}^*}^3}{\pi^2} \frac{1}{2} \left(x \sqrt{1+x^2} - \mathrm{ln}(x+\sqrt{1+x^2})\right)\, , \\
\label{eq:exp_value_v}
     V_0 &= \frac{g_v}{m_v^2} \langle \Psi^\dagger \Psi \rangle = \frac{g_v}{m_v^2} \frac{{M_\mathrm{D}^*}^3}{3 \pi^2} x^3\, ,
\end{align}

where $x=k_f/M^*_D$ is the dimensionless Fermi momentum, and the effective mass of the DM fermion reduces due to the scalar coupling as
\begin{equation}
\label{eq:effective_mass_linear}
    M_\mathrm{D}^* = M_\mathrm{D} - g_{s_1} \phi_0.
\end{equation}
The approximations in Eqs.~\eqref{eq:exp_value_phi_linear}, \eqref{eq:exp_value_v} and \eqref{eq:effective_mass_linear} along with the stress-energy tensor ($\alpha$ runs over all the fields) 
\begin{equation}
\label{eq:stress_energy_tensor_in_MF_theory}
    T^{\mu\nu} = \sum_\alpha \frac{\partial \mathcal{L}}{\partial (\partial_\mu \phi_\alpha)} \partial^\nu \phi_\alpha - \mathcal{L} \eta^{\mu\nu},
\end{equation}
allow computing the energy density and the pressure analytically as \cite{Diener:2008bj}
\begin{align}
\label{eq:energy_density_linear}
    \rho_\mathrm{DM} &= \rho_f + \frac{1}{2} m_v^2 V_0^2 + \frac{1}{2} m_s^2 \phi_0^2 \, , \\
\label{eq:pressure_linear}
    P_\mathrm{DM} &= P_f + \frac{1}{2} m_v^2 V_0^2 - \frac{1}{2} m_s^2 \phi_0^2 \, ,
\end{align}
where $\phi_0$ and $V_0$ are given in Eqs.~\eqref{eq:exp_value_phi_linear}--\eqref{eq:exp_value_v} and the first terms in the right-hand side of Eqs.~\eqref{eq:energy_density_linear}--\eqref{eq:pressure_linear} represent the contribution of DM particles treated as a free Fermi gas as \cite{Diener:2008bj}
\begin{align}
\label{eq:free_density_term}
    \rho_f &= \frac{{M_\mathrm{D}^*}^4}{\pi^2} \frac{1}{8} \left( x \sqrt{1+x^2} (1+2x^2) - \mathrm{ln} (x+\sqrt{1+x^2}) \right) \, , \\
\label{eq:free_pressure_term}
    P_f &= \frac{{M_\mathrm{D}^*}^4}{3\pi^2} \frac{1}{8} \left( x \sqrt{1+x^2} (-3+2x^2) + 3 \, \mathrm{ln} (x+\sqrt{1+x^2}) \right) \, .
\end{align}
Equations~\eqref{eq:energy_density_linear}--\eqref{eq:free_pressure_term} represent the DM EoS in the scalar linear coupling scenario for our study. We have five parameters: the three masses $M_\mathrm{D}, m_v, m_s$ and the two couplings $d_v, d_{s_1}$. In the dark sector, the parameters are not experimentally restricted, allowing for the generation of an extraordinarily large number of DANS configurations. A conservative approach would be to freely vary these parameters to explore their effects on the DANS structure. However, to ensure a physically meaningful parameter selection, we perform a Bayesian analysis to identify the set of parameters that best aligns with the available astrophysical data.

\subsubsection{Scalar quadratic coupling}
\label{sec:quadratic_scalar_coupling}

In the \emph{quadratic coupling} scenario, we keep the same interaction term for the dark vector mediator, i.e., the current $g_v \overline{\Psi} \gamma_\mu \Psi V^\mu$. However, in contrast to the linear coupling Lagrangian Eq.~\eqref{eq:linear_lagrangian}, we assume a quadratic interaction of the scalar mediator with the dark fermion of the form $g_{s_2} \phi^2\overline{\Psi} \Psi$ - see Eq.~\eqref{eq:quadratic_lagrangian}. Furthermore, we include a quartic potential $\supset (\lambda/4!) \; \phi(x)^4$, which is essential within the RMF framework to ensure a non-trivial vacuum expectation value for $\phi$. Indeed, the impact of the quadratic scalar coupling is generally suppressed with respect to the linear one and becomes appreciable only in combination with the self-interaction term. There are well-established theoretical motivations to consider scalar DM characterized by a quartic self-interaction, as reviewed in \cite{Magana:2012ph}, in the context of cosmological inflation \cite{Peebles:1999se}, and in models where it constitutes central galactic halos \cite{Lesgourgues:2002hk, Arbey:2003sj}. From the perspective of Effective Field Theory, linear coupling typically plays a dominant role in phenomenology. Nonetheless, the quadratic coupling can contribute to the DM mass. The linear coupling is suppressed in some technically natural models, allowing the quadratic coupling to become dominant. If an additional approximate discrete $Z^\phi_2$ symmetry is present, it protects the quadratic coupling, causing it to dominate over the linear coupling. Models such as the relaxed relaxion \cite{Banerjee:2020kww} and the clockwork mechanism \cite{Farina:2016tgd} are capable of producing dominant quadratic couplings. Therefore, the Lagrangian incorporating the quadratic interaction is given by
\begin{eqnarray}
\label{eq:quadratic_lagrangian}
    \mathcal{L}&=& \overline{\Psi} (x) \left [ \gamma_{\mu} \left( i \partial^{\mu} - g_v V^{\mu}(x) \right) - \left(M_\mathrm{D} - g_{s_2} \phi^2(x) \right ) \right ] \Psi (x) + \nonumber\\
    &+& \frac{1}{2} \big( \partial_{\mu} \phi (x) \partial^{\mu} \phi (x) -m_{s_2}^2 \phi^2(x) \big) - \frac{\lambda}{4!} \phi^4 (x) - \frac{1}{4} V_{\mu\nu}V^{\mu\nu} + \frac{1}{2} m_v^2 V_\mu (x) V^\mu (x) \, ,
\end{eqnarray}
where the quantities have the same meaning as in Eq.~\eqref{eq:linear_lagrangian}. The vector coupling $g_v$ is still given by Eq.~\eqref{eq:coupling_vector}, whereas, similarly to Section~\ref{sec:linear_scalar_coupling}, the scalar coupling $g_{s_2}$ is rescaled as
\begin{equation}
\label{eq:coupling_scalar_quadratic}
    g_{s_2} = \frac{d_{s_2}  M_\mathrm{D}}{\widetilde{M}_p^2}.
\end{equation}
Note that one advantage of this parameterization is that, while $g_{s_1}$ and $g_{s_2}$ have different mass dimensions ($[g_{s_1}] = 0$ and $[g_{s_2}] = -1$), $d_{s_1}$ and $d_{s_2}$ are both dimensionless. We stress that the Lagrangian in Eq.~\eqref{eq:quadratic_lagrangian} with a well-motivated scalar quadratic interaction has not been tested comprehensively when studying DANSs.

Since only the scalar interaction is modified, Eqs.~\eqref{eq:eq_motion_v} and \eqref{eq:eq_motion_psi} continue to hold for $V_\mu(x)$ and $\Psi(x)$, respectively. The equation of motion for $\phi(x)$ instead becomes

\begin{equation}
\label{eq:eq_motion_phi_quadratic}
    \partial_\mu \partial^\mu \phi(x) + m_{s_2}^2 \phi(x) + \frac{\lambda}{6} \phi^3 (x) = 2 g_{s_2} \overline{\Psi} (x) \phi (x) \Psi (x) \, .
\end{equation}
Unlike the linear coupling scenario, where the vacuum expectation value of the scalar field $\phi$ is uniquely defined by Eq.~\eqref{eq:exp_value_phi_linear}, ambiguity arises in the case of quadratic coupling scenario. In detail, the potential

\begin{equation}
\label{eq:quadratic_potential}
    V (\phi) = \left ( \frac{1}{2} m_s^2 - g_{s_2} \langle \overline{\Psi} \Psi \rangle \right ) \phi^2 + \frac{\lambda}{4!} \phi^4 \, ,
\end{equation}

corresponding to the Lagrangian in Eq.~\eqref{eq:quadratic_lagrangian}, may be stabilized by the solutions $\phi = 0$ and $\phi = \pm \phi_0$ depending on the sign of the coefficient of the first term in the right-hand side of Eq.~\eqref{eq:quadratic_potential}. With our choices of $m_s$ and $g_{s_2}$ (see Section~\ref{sec:bayesian_analysis} for details), we can verify that $\frac{1}{2} m_s^2 - g_{s_2} \langle \overline{\Psi} \Psi \rangle < 0 $ for

\begin{equation}
\label{eq:Psibar_Psi}
    \langle \overline{\Psi} \Psi \rangle = \frac{1}{\pi^2} \int_0^{k_f} dk \, \frac{k^2 \, M_\mathrm{D}^*}{\sqrt{k^2 + {M_\mathrm{D}^*}^2}}   \gtrsim 300 \; \mathrm{MeV^3} \, , 
\end{equation}

where the effective mass is now
\begin{equation}
\label{eq:effective_mass_quadratic}
    M_{D}^* = M_\mathrm{D} - g_{s_2} \phi_0^2 \, .
\end{equation}

\begin{figure*}[h]
\includegraphics[width=0.74\textwidth]{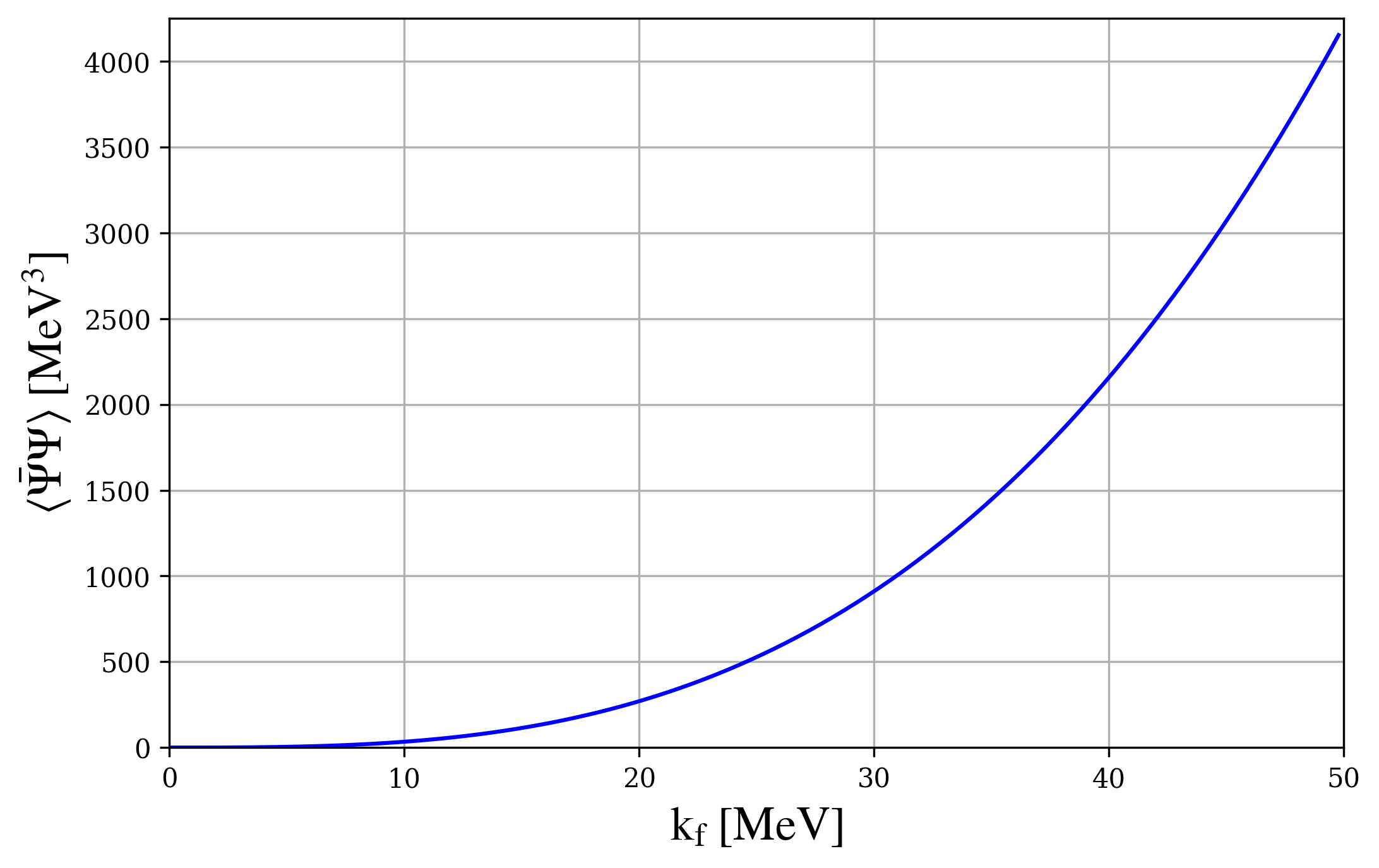}
   \caption{\label{fig:Psibar_Psi_only_large} Variation of $\langle \overline{\Psi} \Psi \rangle$ as function of the Fermi momentum $k_f$ of the DM particle. $\langle \overline{\Psi} \Psi \rangle \gtrsim 300~\mathrm{MeV^3}$ as $k_f \gtrsim 20$ MeV, meaning that our ground state is $\phi_0$.}
\end{figure*}

In FIG.~\ref{fig:Psibar_Psi_only_large} we illustrate the dependence of the scalar density $\langle \overline{\Psi} \Psi \rangle$ of the DM fermions on the Fermi momentum $k_f$. The figure shows that $\langle \overline{\Psi} \Psi \rangle$ exceeds the threshold value required by Eq.~\eqref{eq:Psibar_Psi} already for $k_f \gtrsim 20$ MeV. We anticipate that our DM candidate is nucleon-like, with mass $\sim 1$ GeV. As such\footnote{Assuming that nucleons behave as free Fermi gases, their Fermi energy $E_f = \frac{1}{2m} (3\pi\rho)^{1/3}$ (where $\rho$ is the number density of the particle species) results in $k_f^n \simeq 284$ MeV and $k_f^p \simeq 250$ MeV, by assuming $\rho_n \simeq 0.6 \; A$ ($A$ is the atomic mass of the nucleus) and $\rho_p \simeq 0.4 \; A$ for neutrons and protons, respectively. Hence, our nucleon-like DM fermion is expected to have a comparable Fermi momentum.}, Eq.~\eqref{eq:Psibar_Psi} is verified within our framework, and we can conclude that the non-trivial vacuum configuration $\phi_0 \neq 0$ is realized for physically relevant densities inside DANSs. As a consequence, the potential in Eq.~\eqref{eq:quadratic_potential} is minimized when

\begin{equation}
\label{eq:exp_value_phi_quadratic}
    \phi \rightarrow \pm \phi_0 = \pm \sqrt{\frac{6}{\lambda} \left [ 2 g_{s_2} \langle \overline{\Psi} \Psi \rangle - m_{s_2}^2 \right]} \, .
\end{equation}

Without loss of generality, we can select $+ \, \phi_0$ as only even powers of the scalar field appear in the EoS.

We have demonstrated that the condition in Eq.~\eqref{eq:Psibar_Psi} on the Fermi momentum holds for reasonable parameter values with quite a significant margin. Given this evidence, we expect it to remain valid even for moderately different parameter choices. Nevertheless, a dedicated analysis must be carried out for each specific set of parameters.

Note that, while $\phi_0$ differs when the scalar interaction term is modified, the computation of the ground state mean value of the dark vector mediator $V_0$ is the same as Eq.~\eqref{eq:exp_value_v}.

As in Section~\ref{sec:linear_scalar_coupling}, the EoS is computed by assuming the DM inside a NS as a perfect fluid, whose stress-energy tensor is given in Eq.~\eqref{eq:stress_energy_tensor_in_MF_theory}. The quartic term modifies the density and pressure as
\begin{align}
\label{eq:energy_density_quadratic}
    \rho_\mathrm{DM} = \rho_f + \frac{1}{2} m_v^2 V_0^2 + \frac{1}{2} m_{s_2}^2 \phi_0^2 + \frac{\lambda}{4!} \phi_0^4 \, , \\
\label{eq:pressure_quadratic}
    P_\mathrm{DM} = P_f + \frac{1}{2} m_v^2 V_0^2 - \frac{1}{2} m_{s_2}^2 \phi_0^2 - \frac{\lambda}{4!} \phi_0^4 \, ,
\end{align}
where $\phi_0$ is given in Eq.~\eqref{eq:exp_value_phi_quadratic} and the free terms $\rho_f$ and $P_f$ are still expressed in Eqs.~\eqref{eq:free_density_term}--\eqref{eq:free_pressure_term}, respectively.

Compared to the linear scenario, the DM EoS in Eqs.~\eqref{eq:energy_density_quadratic}--\eqref{eq:pressure_quadratic} has the additional parameter $\lambda$, bringing the total number of parameters to six.

\subsubsection{Bayesian inference of dark matter parameters}
\label{sec:bayesian_analysis}

Bayesian analysis \cite{Thrane:2018qnx} is the main technique to infer the probability distribution of unknown parameters by incorporating prior knowledge and empirical observations. The relationship between unknown parameters and data is established through likelihood functions, which quantify the probability of obtaining the observed data given a certain parameter set. By combining the likelihood with prior information, Bayesian inference helps to estimate the parameters after accounting for the data.

\begin{table*}[h]
\centering
\renewcommand{\arraystretch}{1} 
\setlength{\tabcolsep}{4pt} 

\vspace{10pt}

\begin{tabular}{| c  c  c  c |}
\hline

\hline
DM Parameter & Prior & Nuclear EOS & Posterior\\
\hline

$M_D$ [GeV] & $\mathcal{U}(0.8,\,1.2)$ & $\begin{array}{c} \mathrm{BSk22} \\ \mathrm{MPA1} \\ \mathrm{APR4} \end{array}$ & $\begin{array}{c} 0.971^{+0.0683}_{-0.0332} \\ 1^{+0.0374}_{-0.0429} \\ 0.984^{+0.034}_{-0.0482} \end{array}$ \\
\hline

$m_s$ [MeV] & $\mathcal{N}(10^{-22},\,2\times10^{-23})$ & $\begin{array}{c} \mathrm{BSk22} \\ \mathrm{MPA1} \\ \mathrm{APR4} \end{array}$ & $\begin{array}{c} 0.993^{+0.119}_{-0.0838} \; \times 10^{-22} \\ 0.986^{+0.123}_{-0.0809} \; \times 10^{-22} \\ 1.03^{+0.0595}_{-0.105} \; \times 10^{-22} \end{array}$ \\
\hline

$m_v$ [MeV] & $\mathcal{N}(10,\,2)$ & $\begin{array}{c} \mathrm{BSk22} \\ \mathrm{MPA1} \\ \mathrm{APR4} \end{array}$ & $\begin{array}{c} 10^{+0.854}_{-0.534} \\ 10^{+0.465}_{-0.495} \\ 10.2^{+0.394}_{-0.679} \end{array}$ \\
\hline

$d_{s_1}$ & $\mathcal{U}(2.4\times10^{-7},\;1.7\times10^{-6})$ & $\begin{array}{c} \mathrm{BSk22} \\ \mathrm{MPA1} \\ \mathrm{APR4} \end{array}$ & $\begin{array}{c} 1.05^{+0.191}_{-0.194} \; \times 10^{-6} \\ 0.714^{+0.21}_{-0.161} \; \times 10^{-6} \\ 0.82^{+0.207}_{-0.224} \; \times 10^{-6} \end{array}$ \\
\hline

$d_v$ & $\mathcal{U}(2.4 \times10^{17},\;3.4\times10^{17})$ & $\begin{array}{c} \mathrm{BSk22} \\ \mathrm{MPA1} \\ \mathrm{APR4} \end{array}$ & $\begin{array}{c} 2.85^{+0.0533}_{-0.0335} \; \times 10^{17} \\ 2.97^{+0.0585}_{-0.0256} \; \times 10^{17} \\ 2.95^{+0.0316}_{-0.0418} \; \times 10^{17} \end{array}$ \\
\hline

$f_\mathrm{DM}$ & $\mathcal{U}(0,\,0.25)$ & $\begin{array}{c} \mathrm{BSk22} \\ \mathrm{MPA1} \\ \mathrm{APR4} \end{array}$ & $\begin{array}{c} 9.69^{+0.58}_{-1.25} \; \% \\ 11.3^{+0.984}_{-1.23} \; \% \\ 9.63^{+1.18}_{-1.02} \; \% \end{array}$ \\
\hline

\end{tabular}
\caption{\label{table_bayesian_linear} Summary of the Bayesian parameter estimation for the \emph{linear scalar coupling} scenario. We report the DM parameters appearing in Eq.~\eqref{eq:linear_lagrangian}, their assumed priors (either normal or uniform), the nuclear EOS considered in combination with the dark sector, and the central optimized values of the posterior distributions.}
\end{table*}

\begin{figure*}[!h]
\centering
\includegraphics[width=\textwidth]{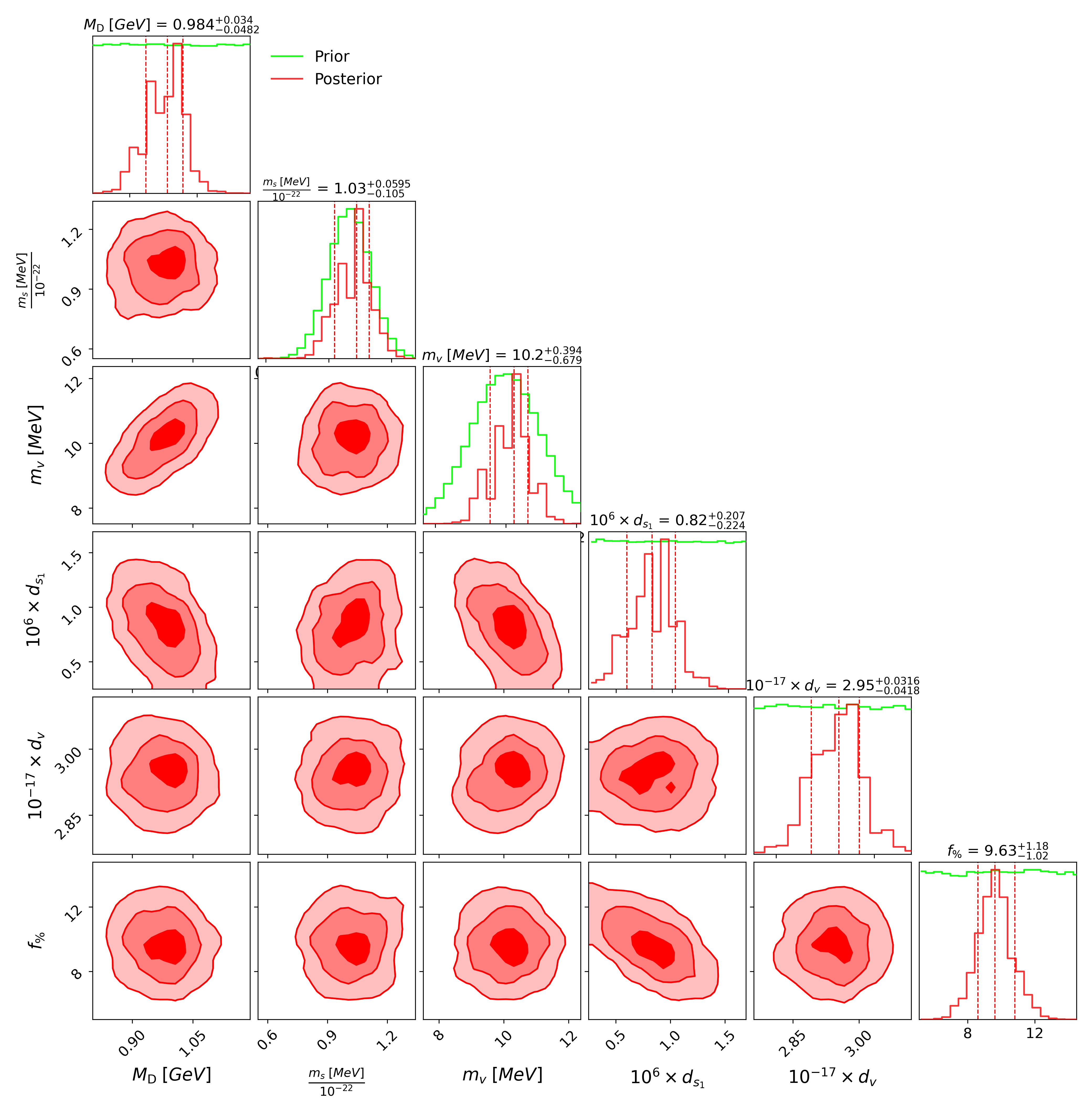}
   \caption{\label{fig:bayesian_linear_APR4} Posterior distributions of the DM parameters in the \emph{linear scalar coupling} scenario when the dark sector is combined with BM described by the APR4 EoS. The contour levels in the corner plots, shaded from lighter to darker red, correspond to the 50\%, 86\%, and 99\% confidence regions; whereas the dashed lines in the 1D marginal distributions represent the 68\% credible interval. No significant correlations arise among the DM parameters.}
\end{figure*}

The application of Bayesian methodology to our case, including the construction of the likelihood function based on Shapiro time delay measurements, NICER observations and BNS merger signals and the numerical framework employed to perform the analysis are described in Appendix~\ref{appendix_1}.

Here, we discuss our choices of the priors. Indeed, selecting the prior distributions for DM parameters is significantly more challenging than for nuclear parameters \cite{Huang:2023grj}. The reason is that, while nuclear EoSs can be constrained by experimental measurements and theoretical considerations, no analogous a priori values exist for DM. Nevertheless, some guidance can be drawn from previous investigations such as \cite{Xiang:2013xwa}. In particular, beginning with the linear scenario, the DM EoS in Eqs.~\eqref{eq:energy_density_linear}-\eqref{eq:pressure_linear} turns out to depend only on the mass of the DM candidate $M_\mathrm{D}$ and on the ratios $c_{s_1} = g_{s_1} / m_s$, $c_v = g_v / m_v$ between the couplings and the masses of the scalar and vector fields, respectively. As a consequence, the parameter space is effectively reduced to $\vec{\theta} = \{ M_\mathrm{D}, c_{s_1}, c_v, f_\mathrm{DM}\}$, requiring only 4 priors to be assigned.

We follow \cite{Das:2020ecp}, which have assumed $c_{s_1} \in [1-7] \; \mathrm{GeV^{-1}} $ and $c_v \in [10-14] \; \mathrm{GeV^{-1}} $. We keep these bounds, as they help narrow down the otherwise wide DM parameter space. However, we assign separate priors to each parameter. 
While this may seem redundant in this context, we anticipate its necessity in the quadratic scenario, where the EoS in Eqs.~\eqref{eq:energy_density_quadratic}-\eqref{eq:pressure_quadratic} still depends on $c_v$ and explicitly on the individual scalar coupling $d_{s_2}$ and mass $m_s$.

In the following, we describe our prior choices, referring to TABLE~\ref{table_bayesian_linear} for a schematic summary.
\begin{itemize}
    \item We select a nucleon-like DM candidate, a widely adopted assumption in the literature \cite{Kain:2021hpk, Karkevandi:2021ygv, Giangrandi:2022wht}. This choice\footnote{Furthermore, it is consistent with the constraint defined in Eq.~(29) of Ref.~\cite{Rutherford:2022xeb}, which restricts the DM candidate mass to a physically motivated range derived from stability and gravitational binding arguments.} is a standard assumption in studies of DANSs, motivated by its potential connection to mirror DM models \cite{Khlopov:1989fj, Ciarcelluti:2010ji}, and by the fact that a GeV-scale fermionic DM candidate naturally leads to Fermi momenta and densities comparable to those encountered in NS interiors. To reflect this expectation while avoiding an overly informative prior, we adopt a broad uniform prior for the DM candidate mass, $\pi(M_\mathrm{D}) = \mathcal{U}(0.8 \; \mathrm{GeV}, 1.2 \; \mathrm{GeV})$ (where $\mathcal{U}$ stands for the uniform distribution), which brackets the nucleon mass and encompasses a $\sim 20\%$ variation around this scale.
    
    \item The condition $c_{s_1} \in [1-7] \; \mathrm{GeV^{-1}} $ implies that $m_s$ and $d_{s_1}$ are degenerate. To disentangle these parameters, we rely on additional findings in the literature. In detail, we refer to Banerjee et al. \cite{Banerjee:2022sqg}, which examined the effect of an interaction term of the form $\frac{|d_{m_N}|}{\widetilde{M}_p} \, \phi \, m_N \, \overline{\psi_N} \, \psi_N $ between the Dirac field $\psi_N$ of the nucleon and a DM scalar field $\phi$. Their analysis constraints the coupling to be $|d_{m_N}| \lesssim 6\times10^{-6}$. Although this work considered a potential SM-DM interaction, we find it reasonable to extend these constraints to DM-DM interactions for two main reasons. First, the interaction is described by a formally equivalent Lagrangian term, making these bounds a natural starting point. Second, given the significant challenges in directly probing interactions that exclusively involve DM particles, it is pragmatic to adopt analogous constraints that have been rigorously established for SM-DM interactions. Therefore, we select an ultralight DM scalar, whose mass is described by the prior $\pi \left( m_s \right ) = \mathcal{N} (10^{-22} \; \mathrm{MeV},\, 2 \times 10^{-23} \; \mathrm{MeV})$, where $\mathcal{N}$ stands for the normal distribution. Such a distribution corresponds to a flat prior for the coupling, i.e., $\pi \left ( d_{s_1} \right ) = \mathcal{U} \left ( 2.4 \times 10^{-7},\, 1.7 \times 10^{-6} \right )$. We stress that choosing a Gaussian prior for $m_s$ is a practical requirement to lift the degeneracy between $m_s$ and $d_{s_1}$ implied by the constraint on $c_{s_1}$ and to ensure a well-behaved Bayesian inference.
    
     \item A similar approach is adopted for the DM vector, though we have more freedom. As shown in Sections~\ref{sec:linear_scalar_coupling}-\ref{sec:quadratic_scalar_coupling}, since the dark vector appears only through the ratio $c_v$ in both scenarios, the degeneracy between $m_v$ and $d_v$ cannot be broken. As long as the condition $c_v \in [10,\,14] \; \mathrm{GeV^{-1}}$ holds, the EoS and the resulting stellar properties remain independent of the individual parameter choices. Inspired by Diamond et al.  \cite{Diamond:2021ekg}, who demonstrated that in BNS mergers dark photons with masses $\in [1-100]$ MeV can yield luminosities $\gtrsim 10^{46}$ ergs that can be used to probe a large region of the unconstrained DM parameter space, we leverage this freedom to assume a heavy DM vector, whose prior is assumed to be $\pi(m_v) = \mathcal{N} (10 \; \mathrm{MeV};\, 2 \; \mathrm{MeV})$. In analogy with the scalar sector, the choice of a Gaussian prior for $m_v$ follows the same strategy adopted for $m_s$: we encode our preference for a physically motivated mass scale through an informative prior on the mass, while retaining a flat prior for the corresponding coupling. The choice $m_v >> m_s$ is further motivated to obtain a potential given in Eq.~\eqref{eq:yukawa_potential}, which is attractive at large distances and repulsive at short distances. Finally, the condition on $c_v$ consequently implies $\pi (d_v) = \mathcal{U} [2.4 \times 10^{17},\, 3.4 \times 10^{17}]$. 
    The coupling $d_v$ may appear too large. However, we emphasize again that in this case our choices are purely conventional. At the same time, we also stress that, unlike previous studies, our numerical infrastructure allows us to independently define the prior for each parameter and compute their posterior separately.
    
    \item The last parameter to be set is the DM fraction $f_\mathrm{DM}$. Some studies \cite{Karkevandi:2021ygv} (with $f_\mathrm{DM} \in [10-50]\%$) or \cite{Giangrandi:2022wht} (even up to 99\%) explored scenarios in which a large amount of DM settles inside the NS. However, smaller fractions are generally expected \cite{Rutherford:2022xeb}. Therefore, we assume the same prior in \cite{Das:2020ecp}, i.e., $\pi(f_\mathrm{DM}) = \mathcal{U}(0,\,0.25)$.   
\end{itemize}

\newpage 

\begin{table*}[h]
\centering
\renewcommand{\arraystretch}{1} 
\setlength{\tabcolsep}{4pt} 

\vspace{10pt}

\begin{tabular}{| c  c  c  c |}
\hline

\hline
DM Parameter & Prior & Nuclear EOS & Posterior\\
\hline

$M_D$ [GeV] & $\mathcal{\mathcal{U}}(0.8,\,1.2)$ & $\begin{array}{c} \mathrm{BSk22} \\ \mathrm{MPA1} \\ \mathrm{APR4} \end{array}$ & $\begin{array}{c} 1.02^{+0.0371}_{-0.0434} \\ 1.03^{+0.0381}_{-0.0431} \\ 0.982^{+0.0529}_{-0.031} \end{array}$ \\
\hline

$m_s$ [eV] & $\mathcal{N}(10^{-22},\,2\times10^{-23})$ & $\begin{array}{c} \mathrm{BSk22} \\ \mathrm{MPA1} \\ \mathrm{APR4} \end{array}$ & $\begin{array}{c} 1.01^{+0.138}_{-0.122} \; \times 10^{-22} \\ 1.02^{+0.132}_{-0.137} \; \times 10^{-22} \\ 1.03^{+0.147}_{-0.132} \; \times 10^{-22} \end{array}$ \\
\hline

$m_v$ [MeV] & $\mathcal{N}(10,\,2)$ & $\begin{array}{c} \mathrm{BSk22} \\ \mathrm{MPA1} \\ \mathrm{APR4} \end{array}$ & $\begin{array}{c} 10^{+0.557}_{-0.424} \\ 10.1^{+0.504}_{-0.508} \\ 9.93^{+0.695}_{-0.464} \end{array}$ \\
\hline

$d_{s_2}$ & $\mathcal{U}(2.4\times10^{-7},\;1.7\times10^{-6})$ & $\begin{array}{c} \mathrm{BSk22} \\ \mathrm{MPA1} \\ \mathrm{APR4} \end{array}$ & $\begin{array}{c} 0.748^{+0.578}_{-0.342} \; \times 10^{-6} \\ 0.838^{+0.463}_{-0.418} \; \times 10^{-6} \\ 0.824^{+0.54}_{-0.388} \; \times 10^{-6} \end{array}$ \\
\hline

$d_v$ & $\mathcal{U}(2.4 \times10^{17},\;3.4\times10^{17})$ & $\begin{array}{c} \mathrm{BSk22} \\ \mathrm{MPA1} \\ \mathrm{APR4} \end{array}$ & $\begin{array}{c} 2.63^{+0.06}_{-0.0529} \; \times 10^{17} \\ 2.68^{+0.0638}_{-0.0596} \; \times 10^{17} \\ 2.7^{+0.0624}_{-0.0577} \; \times 10^{17} \end{array}$ \\
\hline

$\lambda$ & $\mathcal{U}(5\times10^{-33},\,1\times10^{-31})$ & $\begin{array}{c} \mathrm{BSk22} \\ \mathrm{MPA1} \\ \mathrm{APR4} \end{array}$ & $\begin{array}{c} 6.48^{+2.44}_{-2.2} \; \times 10^{-32} \\ 6.51^{+2.36}_{-2.69} \; \times 10^{-32} \\ 6.53^{+2.19}_{-2.46} \; \times 10^{-32} \end{array}$\\
\hline

$f_\mathrm{DM}$ & $\mathcal{U}(0,\,0.25)$ & $\begin{array}{c} \mathrm{BSk22} \\ \mathrm{MPA1} \\ \mathrm{APR4} \end{array}$ & $\begin{array}{c} 11.1^{+0.604}_{-0.529} \; \% \\ 12.5^{+0.839}_{-0.705} \; \% \\ 10.7^{+0.69}_{-0.758} \; \% \end{array}$ \\
\hline

\end{tabular}
\caption{\label{table_bayesian_quadratic} Summary of the Bayesian parameter estimation for the \emph{quadratic scalar coupling} scenario. We report the DM parameters appearing in the Lagrangian \eqref{eq:quadratic_lagrangian}, their assumed priors (either normal or uniform), the nuclear EOS considered in combination with the dark sector, and the central optimized values of the posterior distributions.}
\end{table*}

We present our analysis\footnote{Note that the DM fraction $f_\mathrm{DM}$ is expressed as a percentage $f_\%$ in all plots.} in FIG.~\ref{fig:bayesian_linear_APR4}, where DM is coupled to a representative BM EoS as APR4. In the one-dimensional panels, the green curves trace the sampled priors, allowing a direct comparison with the corresponding posterior distributions. For brevity, the same results obtained with BSk22 and MPA1 are provided in FIGs.~\ref{fig:bayesian_linear_BSk22}, \ref{fig:bayesian_linear_MPA1} in Appendix~\ref{appendix_2}. For each case, we report the central value of the posterior distribution along with their $1 \sigma$ credible intervals. Similarly, the corner plots show three contours that, from lightest to darkest, enclose the 50\%, 86\%, and 99\% credible regions respectively. The corner plots reveal that the DM parameters are essentially uncorrelated, and the one-dimensional posteriors are approximately Gaussian. Our findings are therefore consistent with those in \cite{Das:2020ecp}.

Furthermore, we observe that the different nuclear EoSs do not introduce significant variations in the DM parameters. However, the greater stiffness of MPA1 (see FIG.~\ref{fig:nuclear_EoSs}) leads to a larger central value in the posterior distribution of $f_\mathrm{DM}$. At given mass, MPA1 yields more compact stars with smaller radii than BSk22 and APR4. If the baryonic sector exerts more pressure, a larger amount of (pressureless) DM can accumulate while maintaining the hydrostatic equilibrium, making higher DM fractions more probable. We also note a mild EoS dependence in the scalar and vector couplings, with $d_{s_1}$ ($d_v$) slightly smaller (larger) for MPA1. These shifts are, however, modest and remain compatible within the $1\sigma$ credible intervals.

We adopt the same framework in the quadratic coupling scenario, with a few differences (see TABLE~\ref{table_bayesian_quadratic}).
\begin{itemize}
    \item The EoS in Eqs.~\eqref{eq:energy_density_quadratic}-\eqref{eq:pressure_quadratic} still depends on $c_v$, but the contribution of the scalar field is no longer limited to $c_{s_2}$. While no specific condition needs to be imposed on $c_{s_2}$, we keep the same priors on $m_s$ and $d_{s_2}$ for consistency.
    
    \item The self-coupling $\lambda$ is an additional parameter in the DM EoS that has to be evaluated. However, to the best of our knowledge, no prior constraints exist on $\lambda$, making the selection of its prior challenging. To narrow down the possibilities, we note that if we assume that the other DM parameters are the same as in the linear scenario, then values of $\lambda \gtrsim 10^{-32}$ do not provide significant effects on the stellar structure. On the other side, we run into an unphysical regime with pressure $P<0$ when $\lambda \lesssim 10^{-34}$. Given this evidence, we select $\pi(\lambda) = \mathcal{U}(5\times10^{-33},\,1\times10^{-31})$.  
\end{itemize}

\begin{figure*}[h]
\centering
\includegraphics[width=\textwidth]{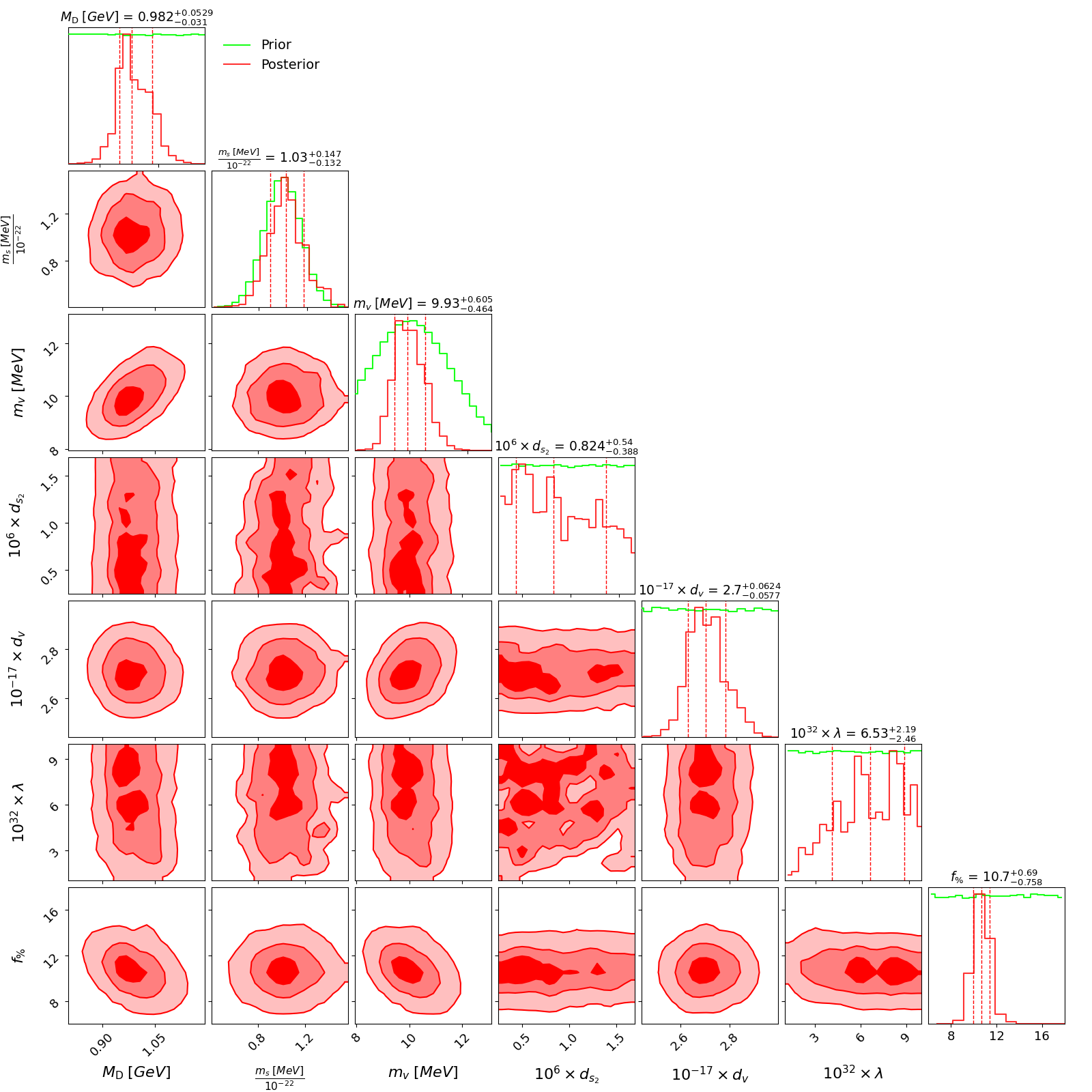}
   \caption{\label{fig:bayesian_quadratic_APR4} Posterior distributions of the DM parameters in the \emph{quadratic scalar coupling} scenario when the dark sector is combined with BM described by the APR4 EoS. The experimental data and the credible regions/intervals are the same of FIG.~\ref{fig:bayesian_linear_APR4}.}
\end{figure*}

The resulting posterior distributions and the corresponding corner plots are displayed in FIG.~\ref{fig:bayesian_quadratic_APR4} for the representative nuclear EoS APR4, while the same outcomes for the other two EoSs considered i.e., BSk22 and MPA1 that are are reported in FIGs.~\ref{fig:bayesian_quadratic_BSk22}, \ref{fig:bayesian_quadratic_MPA1} in Appendix~\ref{appendix_2}. As in the linear scenario, we verify that the parameters are not correlated and that MPA1 leads to a larger $f_\mathrm{DM}$. In the quadratic scenario, where the attractive scalar interaction is strongly suppressed by construction, the Bayesian analysis exhibits a few differences with respect to the linear case. First, the posteriors favor slightly smaller values of the couplings, together with larger DM fractions and a somewhat higher preferred DM mass. This trend is consistent with the fact that, once the net attraction is reduced, a larger amount of DM can be accreted in the star. Moreover, the posterior distribution of $d_{s_2}$ is rather flat as this parameter has a limited impact on the EoS. Finally, the posterior distribution of $\lambda$ is strongly suppressed for $\lambda \lesssim 10^{-33}$, where the corresponding stellar configurations become highly disfavoured, while it approaches a plateau for larger values.

\subsubsection{Dark matter sound speed}
\label{sec:DM_sound_speed}

\begin{figure*} [!ht]
    \includegraphics[width=1\textwidth]{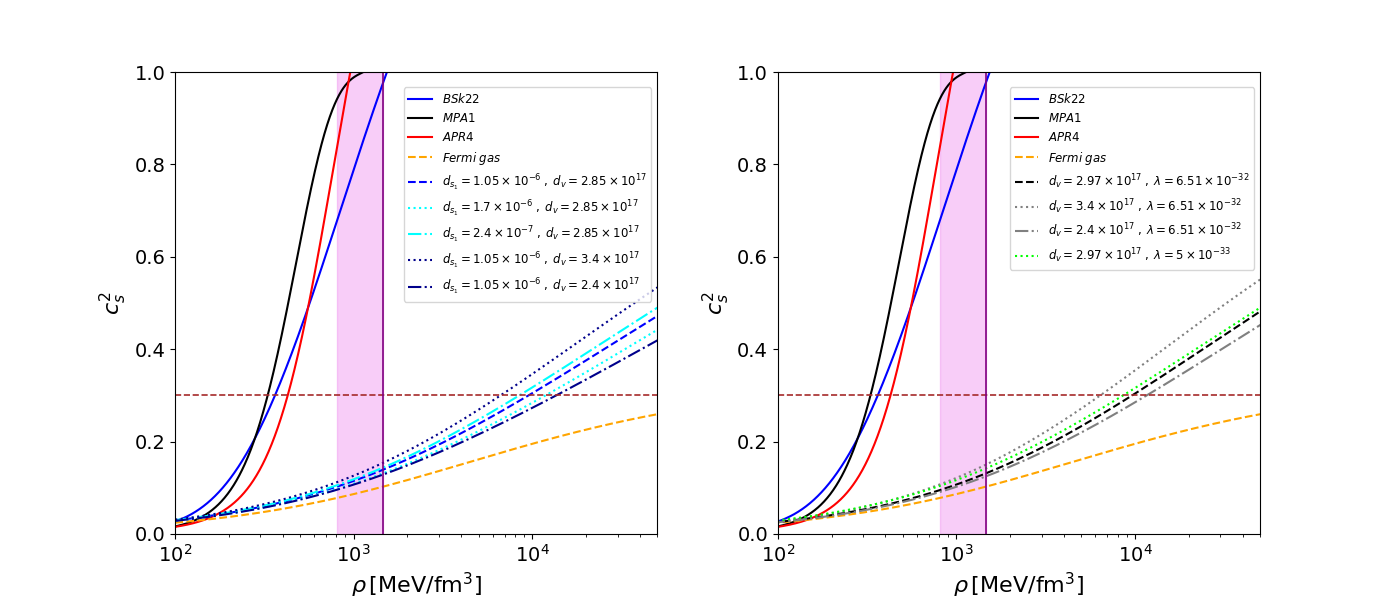}
    \caption{\label{fig:sound_speed} The sound speed $c_s^2$ as a function of energy density $\rho$ for both the baryonic and the dark fluids in the linear (left panel) and in the quadratic (right panel) scenarios. The purple vertical strip represents the 95\% range estimation of the larger central density inside an NS. The brown dashed line denotes the conformal limit $c_s^2/3$, while cyan/dark blue curves represent the same EoS when varying the scalar/vector coupling, while gray/green curves represent the same EoS when varying the vector/$\lambda$ coupling. The details are given in the main text.}
\end{figure*}

To characterize the EoSs that we adopt in our study, we calculate the adiabatic sound speed \cite{Rezzolla:2013dea, Ecker:2022xxj}. For a generic fluid of pressure $P$ and energy density $\rho$, the sound speed (at constant entropy per baryon $S$) is expressed as
\begin{equation}
\label{eq:sound_speed}
    c_s^2 := \left ( \frac{\partial P}{\partial \rho} \right)_S \, .
\end{equation}

While a few studies \cite{Giangrandi:2022wht, Thakur:2024btu} have attempted to generalize the definition Eq.~\eqref{eq:sound_speed} to a whole DANS, in this context we focus on evaluating $c_s^2$ for each component separately. In particular, we aim to first validate our DM EoSs by showing that they do not violate causality (i.e., $c_s^2 < c^2=1$ in all our models, where $c$ is the speed of light); then we analyze the influence of the couplings $d_i$ on the DM sound speed.

In the left panel of FIG.~\ref{fig:sound_speed}, we illustrate the variation of the sound speed with energy density in the linear scalar coupling scenario. The purple band marks the 95\% confidence range for the maximum central energy density inside a NS \cite{Altiparmak:2022bke}. The dashed brown line denotes the QCD (quantum chromodynamics) conformal limit $c^2/3$ representing the upper limit of $c_s^2$ inside a NS at large densities. This limit emerges as QCD's conformal symmetry, restored in this regime \cite{Rezzolla:2013dea}, causing the sound speed to approach the characteristic value $c^2/3$ of conformal field theory realized in ultrarelativistic fluids (whose EoS takes the simplified form $P = \rho / 3$). At lower densities, some studies \cite{Bedaque:2014sqa,Roy:2022nwy} (and references therein) argued that the conformal limit may not be strictly valid due to the non-relativistic nature of particles\footnote{This would also be the case of cold DM.}. The solid curves approaching $c^2$ in the core of a NS depict the sound speed of nuclear matter as described by BSk22 (blue), MPA1 (black), APR4 (red). Starting from intermediate densities, they all violate the conformal limit \cite{Lin:2021ijx}, confirming the ongoing debate on whether this limit holds.

The sound speeds for the DM EoSs are instead reported with non-solid lines. In particular, we consider a representative\footnote{Since we compute the sound speed for a single fluid, $f_\mathrm{DM}$ is not needed. The other DM parameters are rather similar regardless of the nuclear sector (TABLE~\ref{table_bayesian_linear}).} DM EoS (BSk22, with dashed blue line) by selecting the central values of the posterior distributions in FIG.~\ref{fig:bayesian_linear_BSk22}. Although this choice corresponds to the most likely scenario based on Bayesian inference, we also explore the effect of varying couplings\footnote{In principle we could also vary the masses. However, we assume the masses of the DM fields follow Gaussian distributions, making significant deviation from their central values unlikely. Furthermore, our focus is on assessing the role of DM interactions in shaping compact star properties.}. In detail, we examine how the sound speed modifies when either $d_{s_1}$ or $d_v$ is set to its maximum (dotted lines) or minimum (dash-dotted curves) value within the selected priors (TABLE~\ref{table_bayesian_linear}). The resulting EoSs are depicted in cyan and dark blue for variations in the scalar and vector coupling, respectively. We observe that a larger $d_v$ and/or a smaller $d_{s_1}$ increases $c_s^2$. This is expected, as the vector field mediates a repulsive interaction. Increasing its strength raises the pressure, making the star harder to compress and thus stiffening the EoS. Conversely, the scalar field mediates an attractive force that soften the EoS: reducing $d_{s_1}$ has, therefore, the same qualitative effect. On the other hand, $c_s^2$ decreases with a greater attractive interaction, namely with smaller $d_v$ and/or larger $d_{s_1}$.

Although the couplings vary within narrow intervals, even modest changes produce discernible effects on the sound speed. These variations are more pronounced when the vector coupling $d_v$ deviates from its central value, highlighting the dominant role of the repulsive interaction in our framework. Moreover, the influence of the DM interactions on $c_s^2$ is enhanced in the high-density DANS core ($\rho \gtrsim 10^{3} \; \mathrm{MeV/fm^3}$), while remaining relatively weak in the outer layers.

In contrast to the nuclear EoSs, all DM models satisfy the conformal limit at densities typically reached inside a NS. The violation of the limit at very high densities confirms the findings of Das et. al \cite{Das:2021yny} and He et al. \cite{He:2022yrk}, where Lagrangians similar to ours -- Eqs.~\eqref{eq:linear_lagrangian}-\eqref{eq:quadratic_lagrangian} -- provide sound speeds that asymptotically approach $c$. While the agreement with the conformal limit inside NSs might stem from the relative simplicity of our model, a more profound assessment requires computing $c_s^2$ of the whole DANS, which is beyond the scope of our present work.

Finally, we also show the case where both interactions are ``switched off'' ($d_v = d_{s_1} = 0$). Our EoS in Eqs.~\eqref{eq:energy_density_linear}-\eqref{eq:pressure_linear} reduces to a free Fermi gas description, whose sound speed asymptotically approaches $1/3$ \cite{Zhou:2023ndi}. Since the repulsive interaction dominates over the attractive one, turning it on leads to a significant increase in $c_s^2$.
 
The DM sound speed in the quadratic scalar coupling scenario is shown in the right panel of FIG.~\ref{fig:sound_speed}. Here, we select the central values of the posterior distributions in FIG.~\ref{fig:bayesian_quadratic_MPA1} to evaluate our representative DM EoS (MPA1, with dashed black curve). Then, to asses the roles of interactions, we vary one coupling while keeping all other parameters fixed. The effect of changing $d_v$ follows the same pattern: the dotted (dashed) gray curve indicates that a higher (lower) $d_v$ leads to a larger (lower) $c_s^2$ as a consequence of the increasing (decreasing) repulsion. On the other hand, varying $d_{s_2}$ within the range $[2.4\times10^{-7},\; 1.7\times10^{-6}]$ does not significantly affect $c_s^2$. Such a behavior is expected since the scalar field expectation value in Eq.~\eqref{eq:exp_value_phi_quadratic} is suppressed by a factor $\widetilde{M}_p^2$ compared to the linear scenario -- Eqs.~\eqref{eq:coupling_vector}-\eqref{eq:exp_value_phi_linear}. However, from Eq.~\eqref{eq:exp_value_phi_quadratic}, a more pronounced effect of the DM scalar field may still arise if $m_s^2/\lambda \sim \mathcal{O}(1) \; \mathrm{MeV^2}$, which, with our chosen priors, occurs at $\lambda \sim 5 \times 10^{-33}$ (green dashed line). In this regime, the scalar term $\propto m_{s_2}^2 \phi_0^2$ in Eqs.~\eqref{eq:energy_density_quadratic}-~\eqref{eq:pressure_quadratic} may become increasingly relevant.

\section{Mass-radius and tidal deformability-mass relations}
\label{sec:mrtd}

In this section, we obtain the mass-radius and the tidal deformability-mass relations for the DANS configurations by solving the two fluid TOV equations (Eq.~\eqref{eq:TOV}-\eqref{eq:enclosed_mass}). We analyze both linear and quadratic scalar interactions (in Sections~\ref{sec:linear_scenario_mrtm} and \ref{sec:quadratic_scenario_mrtm}, respectively), first using parameters from the Bayesian analysis in Section~\ref{sec:bayesian_analysis}, then varying one coupling strength as in Section~\ref{sec:DM_sound_speed} to assess its impact.

Depending on the choice of the parameters, the DM can either settle in the core of the hosting star or develop a halo around it. Both configurations may alternate along the entire M-R curve. {To distinguish these cases, we adopt the standard criterion outlined in \cite{Kain:2021hpk}, stating that DANSs feature a DM core if $R_\mathrm{DM} < R_\mathrm{BM}$, or a halo when $R_\mathrm{DM} > R_\mathrm{BM}$. In the former case $R_\mathrm{DANS} \equiv R_\mathrm{BM}$ because DM remains entirely within the star; in the latter situation $R_\mathrm{DANS} \equiv R_\mathrm{DM}$, since DM extends beyond the ``baryonic region''. We note, however, that in both scenarios the visible radius still remains $R_\mathrm{BM}$, which complicates comparison between theoretical predictions and experimental data when a DM halo forms.

\subsection{Linear scenario}
\label{sec:linear_scenario_mrtm}

In the linear scenario we derive DANS configurations by coupling the DM EoS in Eqs.~\eqref{eq:energy_density_linear}-\eqref{eq:pressure_linear} with the three nuclear EoSs. In FIG.~\ref{fig:first_plot} we present the M-R and $\Lambda$-M relations for the compact stars in absence (solid curves) and in presence (dash-dotted curves) of DM. Unstable configurations are instead depicted with dashed and dotted lines, respectively. Our findings are compared with the existing theoretical and experimental constraints. The shaded areas denote observationally allowed intervals, except for the yellow and green regions in the upper left corner that are excluded by the GR constraint $R > 2M $ and the causality condition $R > 3M$ \cite{Paerels:2009pz, Lattimer:2012nd}, respectively. In detail, we include the bounds on the mass and the radius obtained from the 2019 NICER data. The magenta and purple regions represent the 2019 NICER data of PSR J0030+0451 \cite{Miller:2019cac, Riley:2019yda}; the dark and light blue regions are derived from the 2021 NICER data and XMM-Newton observations of PSR J0740+6620 \cite{Miller:2021qha, Riley:2021pdl}. Constraints from NS-NS merger events are also shown. The upper blue strip has been established from GW190814 \cite{LIGOScientific:2020zkf}, while the filled orange (heavier NS) and gray (lighter NS) regions correspond to the estimates from GW170817 of the masses of the two NSs involved in the coalescence \cite{LIGOScientific:2018cki}. In the $\Lambda$-M plots, the gray band represents the $90\%$ confidence upper bound from GW170817 \cite{LIGOScientific:2017vwq}; the light and dark blue regions indicate the same bounds for the primary and secondary compact objects, assuming low-spin priors for GW190425 \cite{Yang:2022ees, LIGOScientific:2020aai}.

\begin{figure*} [!ht]
    \includegraphics[width=1\textwidth]{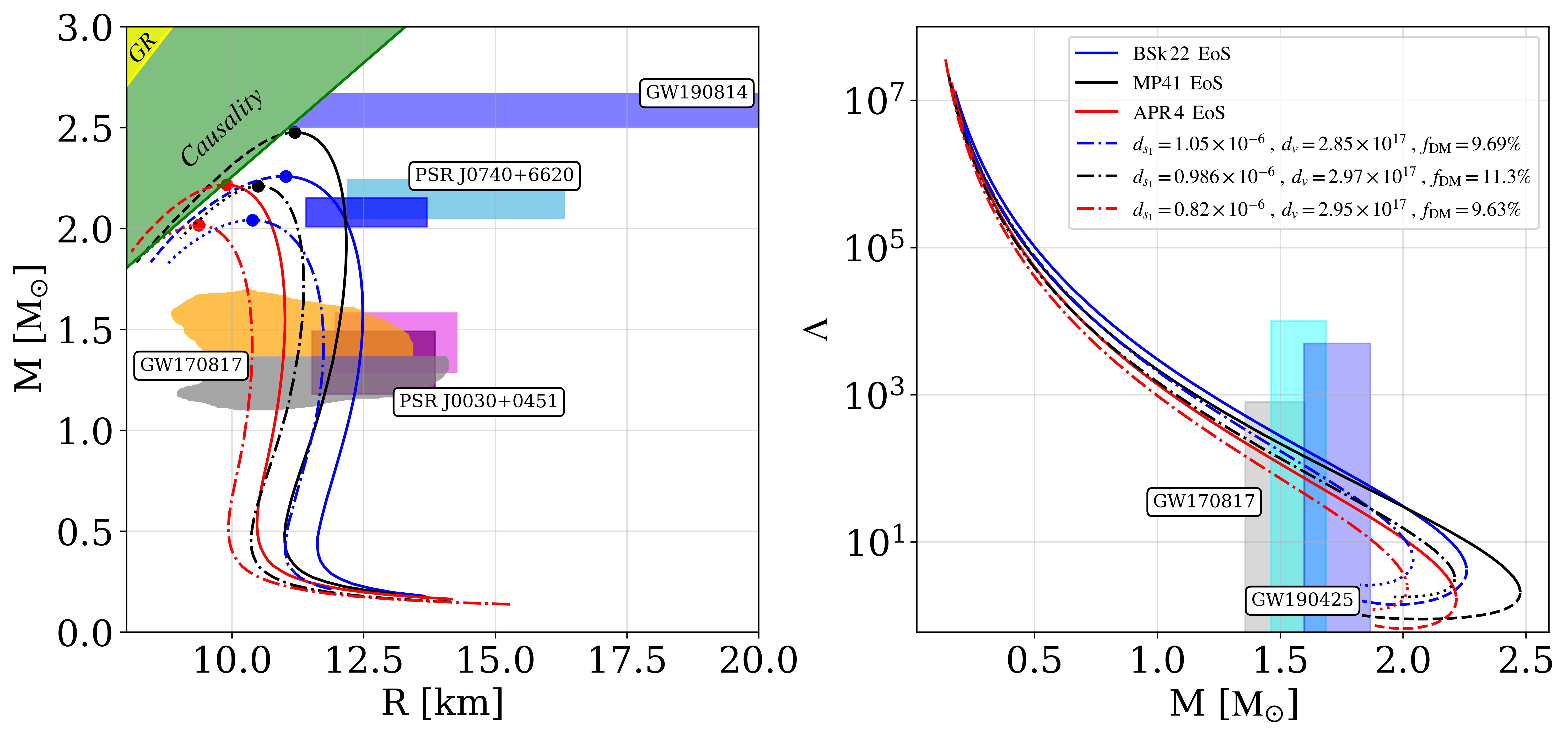}
    \caption{\label{fig:first_plot} Mass-radius (left panel) and tidal-mass (right panel) relations for compact stars with (dash-dotted lines) and without (solid lines) DM. Unstable configurations are instead depicted with dotted and dashed lines, respectively. Experimental constraints are derived from NICER observations as well as GW detections from BNS mergers. The DM parameters to construct the corresponding EoSs have been selected from the Bayesian analysis in Section.~\ref{sec:bayesian_analysis}. DM accumulates in the core of DANSs, which therefore exhibit smaller radii, masses, and tidal deformabilities than pure nuclear EoSs as a consequence of the pressureless nature of the DM.}
\end{figure*}

The DANS configurations are obtained by selecting the DM parameters based on the Bayesian analysis in Section.~\ref{sec:bayesian_analysis}, with each set corresponding to DM coupled with a specific baryonic EoSs. They all exhibit smaller radii, masses and tidal deformabilities compared to those predicted by the respective nuclear EoSs. This is the typical behavior of a DM core being accreted inside a NS \cite{Kain:2021hpk}. As DM accumulates within the stellar volume, the gravitational self-attraction of the star increases. At the same time, because DM does not interact with BM, it does not significantly increase the degeneracy pressure that counteracts the contraction. Although being fermionic, the DM particles interact weakly and so they are not as densely packed as BM in NSs. The resulting degeneracy pressure adds a negligible contribution to the BM. Consequently, the outward pressure can support a lower mass, leading to DANSs that are less massive and more compact. The behavior observed in the right panel of FIG.~\ref{fig:first_plot} reflects the same physical mechanism. For a given stellar mass, the presence of a DM core leads to a smaller stellar radius and, consequently, to a reduced $\Lambda$ with respect to the purely baryonic configurations. As a result, the $\Lambda$-M relations for DANSs are systematically shifted toward lower tidal deformabilities, while remaining compatible with observations.

DANS configurations that develop a DM halo are instead favored by a greater vector coupling $d_v$, as shown in the upper panel of FIG.~\ref{fig:second_third_plot_linear_interactions}. Such configurations are highlighted with yellow shadows along the M-R curves. However, this evidence does not occur when the couplings lie within the range defined by our priors (see TABLE~\ref{table_bayesian_linear}). Rather, we begin to observe DM halos at low central densities with $d_v \sim 4.5 \times 10^{17}$, as demonstrated by the stiffer tail of the M-R relations. Configurations with larger $\rho_c$ still feature DM cores, as the BM dominates gravitational confinement. With even stronger vector coupling ($d_v \sim 6.5 \times 10^{17}$), the repulsion completely overwhelms gravity, so that DM can no longer remain confined within a core regardless of the central density. The increased pressure leads to much larger $R_\mathrm{DANS}$, implying a corresponding increase in the tidal deformability that is computed by means of Eq.~\eqref{eq:Lambda}. However, the maximum TOV mass (which is more sensitive to the DM candidate mass) is not significantly affected as only the strength of the repulsive interaction is varied. This effect is similar across the three nuclear EoSs considered. Our findings are compatible with GW170817, and remain consistent with PSR J0030+0451. On the other hand, the masses of NSs inferred from PSR J0740+6620 and GW190814 appear to exceed what is achievable within our models. This could potentially allow for setting constraints on the abundance and properties of DM, at least when it influences the physics of compact objects. Similar to star mass, the tidal deformability also increases monotonically with a larger vector coupling. All models appear reasonable and could be constrained through NS-NS merger events.  

\begin{figure*} [!ht]
    \includegraphics[width=1\textwidth]{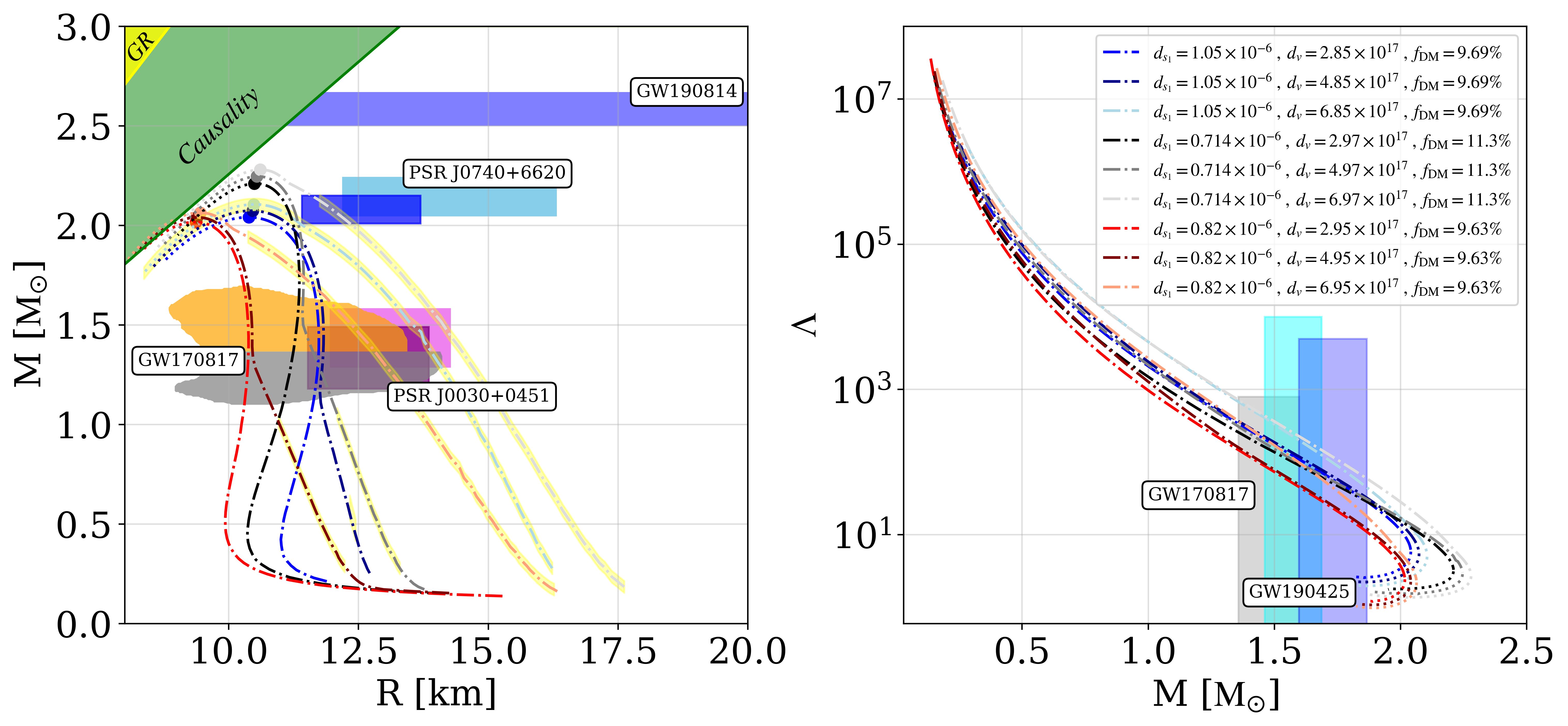}
    \includegraphics[width=1\textwidth]{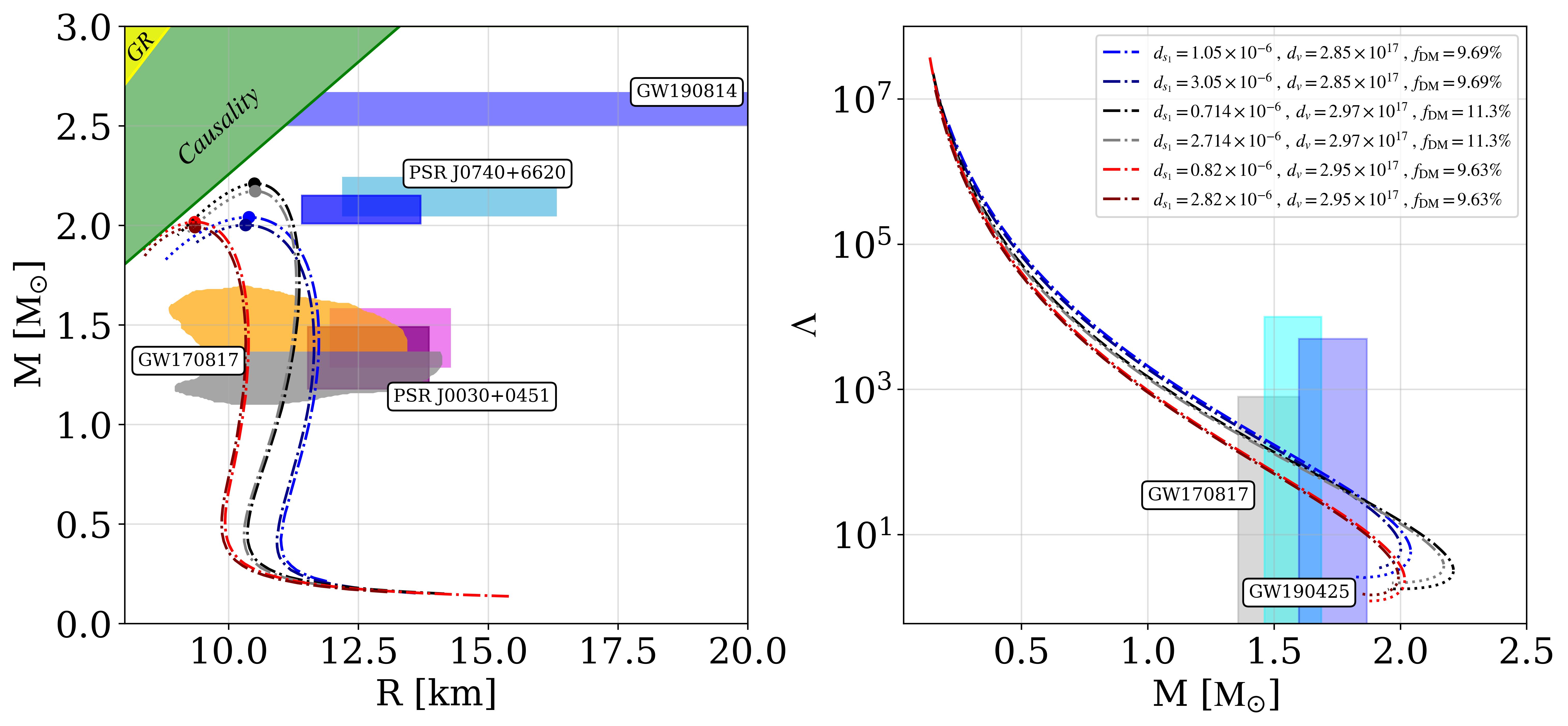}
    \caption{\label{fig:second_third_plot_linear_interactions} Mass-radius (left panel) and tidal-mass (right panel) relations for compact stars with (dash-dotted lines) and without (solid lines) DM.  Unstable configurations are depicted with dotted and dashed lines, respectively. Theoretical and experimental constraints have the same meaning of FIG.~\ref{fig:first_plot}. Upper panel: DANSs configurations obtained by increasing $d_v$. A stronger vector coupling results in larger $R_\mathrm{DANS}$ due to enhanced repulsion, leading to more extended DM halos. Lower panel: DANSs configurations obtained by increasing $d_{s_1}$. Larger scalar coupling corresponds to a more intense attractive force, yielding more compact DM cores and slightly softer EoSs.}
\end{figure*}

The opposite behavior is observed when the scalar coupling $d_{s_1}$ is increased (lower panel in FIG.~\ref{fig:second_third_plot_linear_interactions}). Again, we select values slightly above our flat prior's upper bound to emphasize this effect. As anticipated in the analysis of the DM sound speed in Section.~\ref{sec:DM_sound_speed}, strengthening the attractive interaction softens the EoS describing the DANS, leading to reduced maximum masses and radii. However, the resulting shift in the M-R and $\Lambda$-M relations remains relatively modest, demonstrating that the repulsive force mediated by the vector coupling is generally stronger and thus dominates the overall structure of DANS. This clarifies why in \cite{Collier:2022cpr} only the repulsion was considered. Naturally, no DM halos are observed in this scenario, as the stronger attraction confines the DM entirely within the NS core.

\subsection{Quadratic scenario}
\label{sec:quadratic_scenario_mrtm}

Similar to the previous section, we derive DANS configurations by coupling the DM EoS in Eqs.~\eqref{eq:energy_density_quadratic}-\eqref{eq:pressure_quadratic} with the three mentioned nuclear EoSs. While increasing the strength of the repulsive interaction yields the same effect on DANSs as observed in the linear scenario, a few interesting features emerge when considering the influence of the scalar field. If $\phi$ respects $Z_2$ symmetry \cite{Banerjee:2022sqg}, the suppression $\phi_0 \propto 1/\widetilde{M}_p^2$ makes the DM EoS relatively insensitive to the strength of the scalar coupling, though with $d_{s_2} \gtrsim 10^{4}$ we encounter an unphysical EoS. Nevertheless, as for the sound speed in FIG.~\ref{fig:sound_speed}, the role of the scalar field may manifest when $m_s^2/\lambda \sim \mathcal{O}(1) \; \mathrm{MeV^2}$.

In FIG.~\ref{fig:scalar_quadratic_scenario}, we show that reducing the self-coupling leads to EoSs of comparable stiffness, while allowing configurations with DM halos to form at low central densities. Interestingly, the combination between DM and BM described by BSk22 EoS is the only case where no DM halo forms, proving that the nuclear sector may significantly influence the resulting stellar structure. Unlike the linear coupling, where the DM EoS depends on $c_{s_1}$ and $c_v$, for the quadratic coupling, the DM EoS depends on $c_v$ and explicitly on the individual coupling $d_{s_2}$ and mass $m_s$ of the scalar mediator.

\begin{figure*} [!ht]
    \includegraphics[width=1\textwidth]{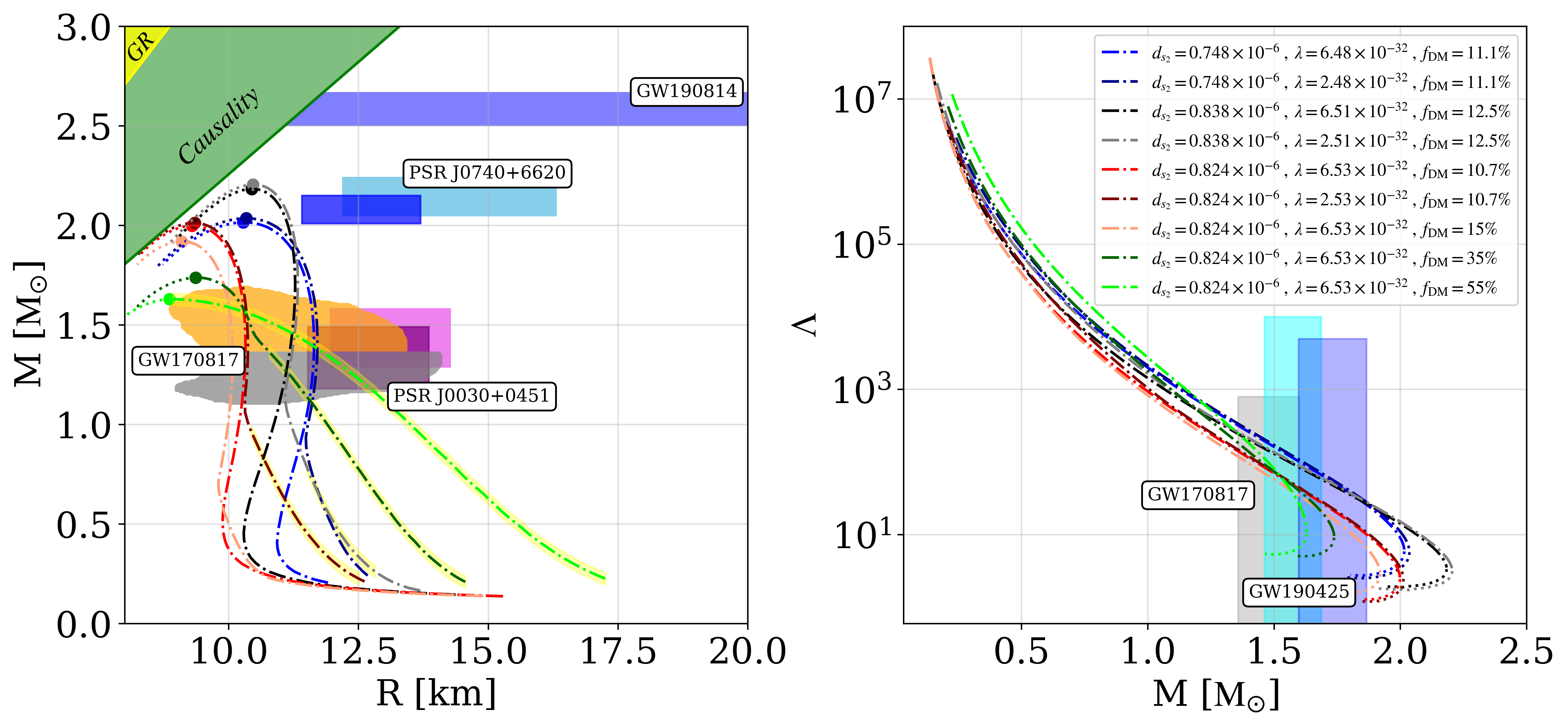}
    \caption{\label{fig:scalar_quadratic_scenario} Mass-radius (left panel) and Tidal-Mass (right panel) relations for compact stars with (dash-dotted lines) and without (solid lines) DM.  Unstable configurations are depicted with dotted and dashed lines, respectively. Theoretical and experimental constraints have the same meaning of FIG.~\ref{fig:first_plot}.
    Reducing the self-interaction strength $\lambda$ favors the formation of DM halos at low central densities.
    Small DM fractions ($f_\mathrm{DM} \sim 15 \%$) yield DANSs with DM cores and reduced $\Lambda$. At moderate fractions ($f_\mathrm{DM} \sim 35 \%$), DM halos begin to form at low $\rho_c$, and become the dominant structure at large fractions ($f_\mathrm{DM} \sim 55 \%$). As such, the tidal deformability increases, exhibiting an overall non-monotonic trend as $f_\mathrm{DM}$ grows.}
\end{figure*}

Furthermore, in FIG.~\ref{fig:scalar_quadratic_scenario} we select APR4 as a representative BM EoS to illustrate the effect of increasing the DM fraction\footnote{Increasing $f_\mathrm{DM}$ has the same qualitative effect in the linear scenario as well as with other nuclear EoSs.} in DANSs. Starting from $f_\mathrm{DM} = 9.5 \%$ (red line), we note that larger concentrations lead to softer EoSs \cite{Grippa:2024ach}, as more DM accumulates without contributing to the overall pressure. For instance, at $f_\mathrm{DM} = 15 \%$ (pink line) the DANS becomes more compact, with a smaller maximum mass, radius, and tidal deformability.

A key feature emerges, however, when $f_\mathrm{DM}$ is increased beyond a certain threshold\footnote{This threshold depends on the specific models adopted for both BM and DM.}. Along the M-R curve (left panel in FIG.~\ref{fig:scalar_quadratic_scenario}), not all DANSs exhibit DM cores. Rather, configurations with lower central densities begin to develop DM halos. In our framework, this core-to-halo transition occurs at $f_\mathrm{DM} \sim 35 \%$ (dark green curve). Remarkably, this transition has a significant impact on the behavior of $\Lambda$. We recall that the tidal deformability is computed through Eq.~\eqref{eq:Lambda}, where $\Lambda \propto (R_\mathrm{DANS} / M_\mathrm{DANS})^5$. As long as DM is concentrated in the core, both $M_\mathrm{DANS}$ and $R_\mathrm{DANS}$ decrease. The reduction in radius dominates \cite{Ivanytskyi:2019wxd}, resulting in an overall decrease of the tidal deformability. However, as soon as a DM halo forms, $\Lambda$ rises due to the combined effect of a still-decreasing $M_\mathrm{DANS}$, and an increasing $R_\mathrm{DANS}$ (that now corresponds to the outermost $R_\mathrm{DM}$).

The alternation between halos and core for moderate DM fractions (e.g., $f_\mathrm{DM} \sim 35 \%$) explains the non-monotonic trend of $\Lambda$ in the right panel in FIG.~\ref{fig:scalar_quadratic_scenario}. In the dark green curve, small central densities lead to DM halos and larger $\Lambda$, whereas greater $\rho_c$ results in DM cores and smaller tidal deformability. Finally, for $f_\mathrm{DM} \gtrsim 0.55$ (light green curve), all stable configurations exhibit DM halos with larger $\Lambda$ compared to DANSs with less DM (red and pink curves). Our findings are consistent with Liu et al. \cite{Liu:2023ecz}, which also found that, despite some dependence on the DM candidate mass, DM halos are typically exclusive for $f_\mathrm{DM} \gtrsim 45 \%$ (see FIGs. 7-8-9 of \cite{Liu:2023ecz}).

Nevertheless, such high values of $f_\mathrm{DM}$ lie well beyond our prior and are unlikely to occur in realistic scenarios \cite{Rutherford:2022xeb}. Moreover, while the visible radius might still be compatible with observations, the fact that the maximum mass reaches only $\sim 1.65 \; M_\odot$ makes it improbable for DANSs to exhibit large DM fractions and be characterized by a dominant halo structure.

\section{Conclusions and discussions}
\label{sec:conclusions}

NSs serve as one of the finest astrophysical laboratories to the search for DM. The direct detection of GWs from LIGO/Virgo, together with high-precision X-ray spectroscopic observations of pulsars, provides a great opportunity to probe new physics. In this paper, we obtain the mass-radius and tidal deformability-mass relations of DANSs, considering both vector and scalar couplings of DM. We also discuss how the stellar properties are affected by linear and quadratic dark scalar couplings with the DM fermions.

We consider three EoSs (BSk22, MPA1 and APR4) of different stiffness for the nuclear sector to evaluate how the gravitational coupling with DM may influence the stellar structure. The DM EoS is a generalization of the free Fermi gas treatment obtained by incorporating attractive and repulsive interactions among dark fermions mediated by a dark scalar and a dark vector field, respectively. In the quadratic scenario, we include a quartic self-interaction $(\lambda/4!) \, \phi^4$ of dark scalars to obtain a non-vanishing expectation value of $\phi_0$. The spontaneous breaking of the $Z_2$ symmetry leads to a degenerate vacuum structure, resulting in spatial regions settling into different vacua and being separated by domain walls. To remain consistent with cosmological observations, we assume that the symmetry breaking occurred prior to inflation, which would dilute any such domain walls. While the phenomenological effects of domain walls, particularly in the context of NS parameter estimation, are indeed an intriguing direction, they lie beyond the scope of the present work. We plan to explore these implications in future studies.

The linear coupling of an ultralight scalar to non-relativistic fermions leads to a Yukawa-type potential, which is tightly constrained by fifth force experiments. In contrast, a quadratic coupling yields a potential that deviates from the Yukawa form and, as a result, is not subject to the same stringent bounds. While both linear and quadratic couplings of dark scalars to SM fermions have been studied in the literature \cite{Banerjee:2022sqg}, we adopt a similar framework for the coupling between dark scalars and dark fermions, motivated by the same considerations.

The unknown DM parameters in our models are determined via the Bayesian parameter optimization technique. Overall, our results favor a nucleon-like dark fermion mass with DM fractions of $\sim \mathcal{O}(10\%)$, which become slightly larger in the quadratic scenario.

More specifically, our main findings in the linear scenario can be summarized as follows:

\begin{itemize}
    \item Unlike Ref.~\cite{Das:2020ecp}, we assign separate priors for the masses $m_i$ and the couplings $d_i$ to establish a consistent framework for the quadratic scenario. The resulting posterior distributions are approximately Gaussian and do not show any significant correlations among the DM parameters.

    According to the Bayesian inference, the preferred configurations feature DM cores, leading to more compact stars with reduced masses, radii, and tidal deformabilities compared to purely baryonic NSs. Moreover, we find evidence that the vector-mediated repulsion dominates the global stellar structure.
\end{itemize}

In the quadratic scenario, we would instead emphasize that:

\begin{itemize}
    \item Such a framework, where a quadratic scalar interaction is supplemented by a quartic self-interaction, has not been systematically investigated for DANSs.

    \item While the scalar coupling $d_{s_2}$ has a reduced impact on the EOS, the suppression of the net attractive interaction allows for larger DM fractions. The resulting stellar properties remain overall compatible with current observational constraints.
\end{itemize}

We also analyze the variations in sound speed of the DM EoS, examining the impact of vector and scalar couplings. The causality condition $c_s^2 < c^2$ is respected across all DANS configurations. This secures the consistency of our approach. The main trends emerging from our analysis point out that:

\begin{itemize}
    \item a stiffer EoS, and consequently a larger $c_s^2$, is observed with larger $d_v$ and smaller $d_{s_1}$/$d_{s_2}$, as the repulsive contribution is increased. Conversely, a stronger/weaker scalar/vector coupling enhances the attraction, leading to slightly softer EoSs. 
    
    \item Within the range of couplings defined by our priors, the DM sound speed satisfies the conformal limit at densities typically reached in NSs, although this limit might be exceeded in denser environments. 
\end{itemize}

A valuable extension of our analysis would involve computing the sound speed across the entire DANS and, consequently, gaining a deeper understanding of how DM affects the potential violation of the conformal limit in NSs. Improving upon these aspects is object of our future work.

As mentioned above, DANS configurations tend to exhibit DM cores. Indeed, when DM is accreted into a NS, the gravitational attraction increases without being balanced by additional pressure.  

We find that increasing the vector coupling $d_v$ yields DANSs that are more prone to developing halos due to the enhanced repulsion. Specifically, for $d_v \gtrsim 6.5 \times 10^{17}$, gravity is completely overwhelmed, resulting in configurations where DM halos form. In contrast, larger couplings $d_{s_i}$ soften the EoS, and more compact DM cores originate. Remarkably, the vector contribution is more pronounced, while the effect of the scalar interaction is typically smaller in the linear scenario and considerably suppressed in the quadratic case. Nevertheless, in the latter scenario, the dark scalar field may still exert a more significant influence if $m_s^2/\lambda \sim \mathcal{O}(1) \; \mathrm{MeV^2}$.

The fraction of DM accumulated within the NS is another free parameter in our models. The Bayesian analysis in Section~\ref{sec:bayesian_analysis} provides posterior distributions' central values $\sim 10$ \%, which result in DANSs with DM cores. As more DM is accreted, the overall EoS becomes softer. For relatively small fractions ($f_\mathrm{DM} \sim 15 \%$), only DM cores are formed. However, at greater fractions, a core-to-halo transition occurs along the M-R curve ($f_\mathrm{DM} \sim 35 \%$), eventually leading to the exclusive formation of DM halos ($f_\mathrm{DM} \sim 55 \%$). As $f_\mathrm{DM}$ increases, the maximum mass progressively decreases, reaching values as low as $\sim 1.65 M_\odot$ at $f_\mathrm{DM} = 55 \%$ (see FIG.~\ref{fig:scalar_quadratic_scenario}). At the same time, the tidal deformability exhibits a non-monotonic trend. For configurations that feature a DM core, $\Lambda$ decreases due to the dominant effect of a reduced stellar radius (corresponding to $R_\mathrm{BM}$). Conversely, in configurations with DM halos, the outermost radius (now corresponding to $R_\mathrm{DM}$) becomes significantly larger, yielding a larger tidal deformability compared to compact stars with no/less DM. While this is an interesting outcome of our analysis, configurations with large DM fractions and halo structures may be in tension with observational data, as many NSs have measured masses $\gtrsim 1.5 M_\odot$.

With the exception of these high-$f_\mathrm{DM}$ configurations, our findings are consistent with the experimental bounds provided by NICER observations and detected NS merger events such as GW170817 in terms of both M-R and $\Lambda$-R relations. In contrast, the lightest object in GW190814 ($M \simeq 2.6 \; M_\odot$) lies well above the maximum mass supported by our DANS configurations, and therefore cannot be accomodated within our model parameter space if interpreted as a NS. Altogether, these results illustrate the power of astrophysical observations to both validate and rule out classes of new particle-sector models through their imprint on compact star properties.

\section*{Acknowledgments}
F. G. thanks C. Huang for useful discussions on the Bayesian analysis. This publication is based on work carried out within the COST Actions COSMIC WISPers (CA21106) and BridgeQG (CA23130), supported by COST (European Cooperation in Science and Technology). T. K. P. would also like to thank the Galileo Galilei Institute for Theoretical Physics for the hospitality and the INFN for partial support during the completion of this work. The authors are indebted to the anonymous referee for helpful suggestions that improved the manuscript.

\appendix
\section{Details on Bayesian analysis}
\label{appendix_1}

In the following, we discuss the methodologies for the Bayesian analysis mentioned in Section~\ref{sec:bayesian_analysis}. Within the Bayesian framework, the posterior distribution $p \left (\vec{\theta}, \vec{x} \right )$ of unknown parameters $\{\theta_1, \theta_2, ..., \theta_n \}$ can be expressed through Bayes' theorem as

\begin{equation}
\label{eq:posterior_distribution}
    p \left( \vec{\theta}; \vec{x} \right ) = \frac{\pi \left ( \vec{\theta} \right ) \times \mathcal{L} \left (\vec{x}; \vec{\theta} \right)}{p \left (\vec{x} \right )} \; ,
\end{equation}

where $\vec{\theta}$ and $\vec{x}$ denote the vectors of parameters and data, respectively; $\pi \left ( \vec{\theta} \right ) $ is the \emph{prior probability distribution}, encoding prior knowledge before considering the data; $\mathcal{L} \left (\vec{x}; \vec{\theta} \right)$ is the likelihood function; and $p(\vec{x})$, known as the \emph{evidence}, serves as a normalization factor in Eq.~\eqref{eq:posterior_distribution}, since it is independent of the parameters $\vec{\theta}$. In our case, we stress that the vector of parameters is nothing but
\begin{align}    
    \label{eq:parameters_linear}
    \vec{\theta_1} &= \{ M_\mathrm{D}, m_v, m_s, d_v, d_{s_1}, f_\mathrm{DM} \} \\
    \label{eq:parameters_quadratic}
    \vec{\theta_2} &= \{ M_\mathrm{D}, m_v, m_s, d_v, d_{s_2}, \lambda, f_\mathrm{DM}\}, 
\end{align}
for the linear and the quadratic scenario, respectively. We recall $f_\mathrm{DM}$ is the fraction of DM accreted into the NSs, as defined in Eq.~\eqref{eq:DM_fraction_definition}.

To obtain the posteriors from Eq.~\eqref{eq:posterior_distribution}, both the prior distributions and the likelihood function must be specified. The selection of the former is comprehensively addressed in Section~\ref{sec:bayesian_analysis}, as it needs careful customization for DM. For the latter, we exploit the factorization of the likelihood after imposing some of the latest experimental bounds on NSs. In particular, we take into account three observational constraints: the mass estimations of known pulsars via Shapiro time delay, the mass and radius measurements from NICER, and the knowledge of mass and tidal deformability from BNS mergers. Given that all the measurements are independent, the total likelihood is given by the product of the likelihoods of single observations:

\begin{equation}
    \label{eq:total_likelihood}
    \mathcal{L} \left( \vec{x}; \vec{\theta} \right) = \prod_{i=1}^{N_i} \mathcal{L}^\mathrm{M} \left( x_i; \vec{\theta} \right) \times \prod_{j=1}^{N_j} \mathcal{L}^\mathrm{NICER} \left( x_j; \vec{\theta} \right) \times \prod_{k=1}^{N_k} \mathcal{L}^\mathrm{GW} \left( x_k; \vec{\theta} \right) \, , 
\end{equation}

having included $N_i$ data from mass measurements, $N_j$ from NICER, $N_k$ from GWs detections into the inference framework.

In our case, $N_i = 3$ because we take into account estimations of masses (at 68.3\% CL) of PSR J0740+6620 as $M = 2.08^{+0.07}_{-0.07} \; M_\odot$ \cite{Fonseca:2021wxt}, for PSR J0348+0432, $M = 2.01^{+0.04}_{-0.04} \; M_\odot$ \cite{Antoniadis:2013pzd}, and for PSR J1614-2230, $M = 1.908^{+0.016}_{-0.016} \; M_\odot$ \cite{NANOGrav:2017wvv}. The likelihood of the $i$-th pulsar is assumed to be Gaussian

\begin{equation}
    \label{eq:likelihood_M}
    \mathcal{L}^M \left ( x_i; \vec{\theta} \right ) = \frac{1}{\sqrt{2 \pi \sigma_i}} \; \mathrm{exp} \left \{-\frac{ \left [M_i - M ( \vec{\theta}) \right ]^2}{2\sigma_i^2}\right \} \, ,
\end{equation}

where $M_i$ and $\sigma_i$ are the mass and the error of such a measurement, respectively; whereas $M(\vec{\theta})$ represents the pulsar mass as predicted by the theoretical model. In other words, $M(\vec{\theta})$ is computed by solving the 2-fluid TOV equations (Eq.~\eqref{eq:TOV}-\eqref{eq:enclosed_mass}) for a given set of parameters, chosen according to the selected model -- either linear Eq.~\eqref{eq:parameters_linear} or quadratic Eq.~\eqref{eq:parameters_quadratic}.

We select two NICER measurements ($N_j = 2$), namely PSR J0030+0451 and PSR J0740+6620. Following\footnote{Note that we define the Bayes' theorem Eq.~\eqref{eq:posterior_distribution} in terms of probability distribution functions rather than total probabilities as in \cite{Jiang:2022tps}.} \cite{Jiang:2022tps}, we employ the fit \href{https://zenodo.org/records/5506838}{\texttt{ST+PST}} for the former and the data file in \href{https://zenodo.org/records/4697625}{\texttt{STU/NICERxXMM/FI\_H/run10}} for the latter. Similarly to \cite{Jiang:2022tps}, the likelihood of th $j$-th pulsar is then estimated via the Kernel Density Estimation (KDE) with a Gaussian kernel, based on the publicly available posterior (M, R) samples $S$, i.e., $\mathcal{L}^\mathrm{NICER} \left ( x_j; \vec{\theta}\right ) \propto \mathrm{KDE} \left ( M(\vec{\theta}), R(\vec{\theta)} \, ; \, S \right )$. For details on the utility and the properties of Gaussian kernels, we refer to \cite{Miller:2019cac, Miller:2021qha} (and references therein).

Finally, $N_k = 2$ as we include data from two merger events, \href{https://dcc.ligo.org/LIGO-P1800370/public/}{GW170817} and \href{https://dcc.ligo.org/LIGO-P2000223/public/}{GW190425}. Despite a small chance to involve a light black hole, we assume only NSs belong to these events. Inspired by \cite{Raaijmakers:2019dks, Raaijmakers:2021uju}, we adopt the low-spin posterior distributions on tidal deformability
and mass ratio with the \texttt{IMRPhenomPv2\_NRTidal12} waveform model. From \cite{LIGOScientific:2018hze, LIGOScientific:2020aai}, the likelihood of each event can be numerically estimated via a KDE on the posterior samples, i.e., $\mathcal{L}^\mathrm{GW} \left ( x_k; \vec{\theta}\right ) \propto \mathrm{KDE} \left ( M(\vec{\theta}), R(\vec{\theta)} \, ; \, S(\vec{\alpha}), W \right )$. Here, the likelihood is evaluated using a KDE based on the posterior samples $S(\vec{\alpha})$, which encode the dependence on the parameters $\vec{\alpha}$ (such as the tidal deformabilities $\Lambda_{1,2}$ of the two stars, the chirp mass $M_c$ and the mass ratio $q$ of the binary) inferred by ground-based interferometers. Finally, the construction of the KDE further depends on the adopted waveform model $W$.

Within our inference framework, the weighted sampling of the parameter vector $\vec{\theta}$ is performed through the nested sampling Monte Carlo algorithm MLFriends \cite{Buchner_2014, Buchner_2019} that uses \texttt{UltraNest} package \cite{buchner2021ultranestrobustgeneral, Huang:2024rfg}. In our analysis, we employ $\sim 10000$ live points for each run to ensure an accurate exploration of the EoS parameter space.

\section{Additional corner plots with different nuclear EoSs}
\label{appendix_2}

For completeness, we include here the additional corner plots that we derive via the Bayesian analysis described in Section.~\ref{sec:bayesian_analysis}. In detail, FIGs.~\ref{fig:bayesian_linear_BSk22}, \ref{fig:bayesian_linear_MPA1} show the posterior distributions of the DM parameters in the \emph{linear scalar coupling} scenario when the dark sector is combined with BM described by BSk22 and MPA1, respectively.
Similarly, FIGs.~\ref{fig:bayesian_quadratic_BSk22}, \ref{fig:bayesian_quadratic_MPA1} show the same corner plots in the \emph{quadratic scalar coupling} scenario.

\begin{figure*}[h]
\centering
\includegraphics[width=\textwidth]{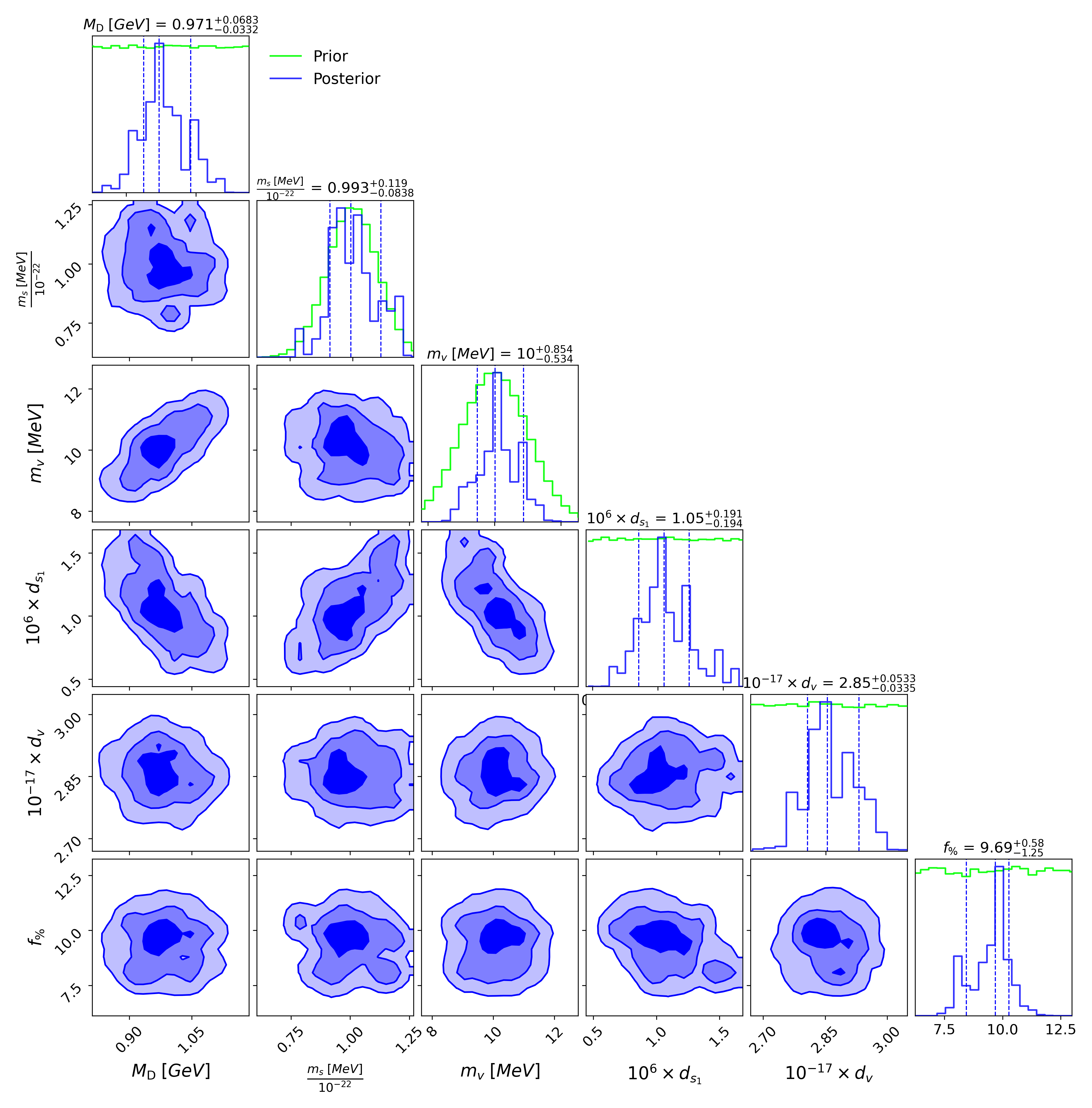}
   \caption{\label{fig:bayesian_linear_BSk22} Same of FIG.~\ref{fig:bayesian_linear_APR4} when DM is combined with BM described by the BSk22 EoS.}
\end{figure*}

\begin{figure*}[h]
\centering
\includegraphics[width=\textwidth]{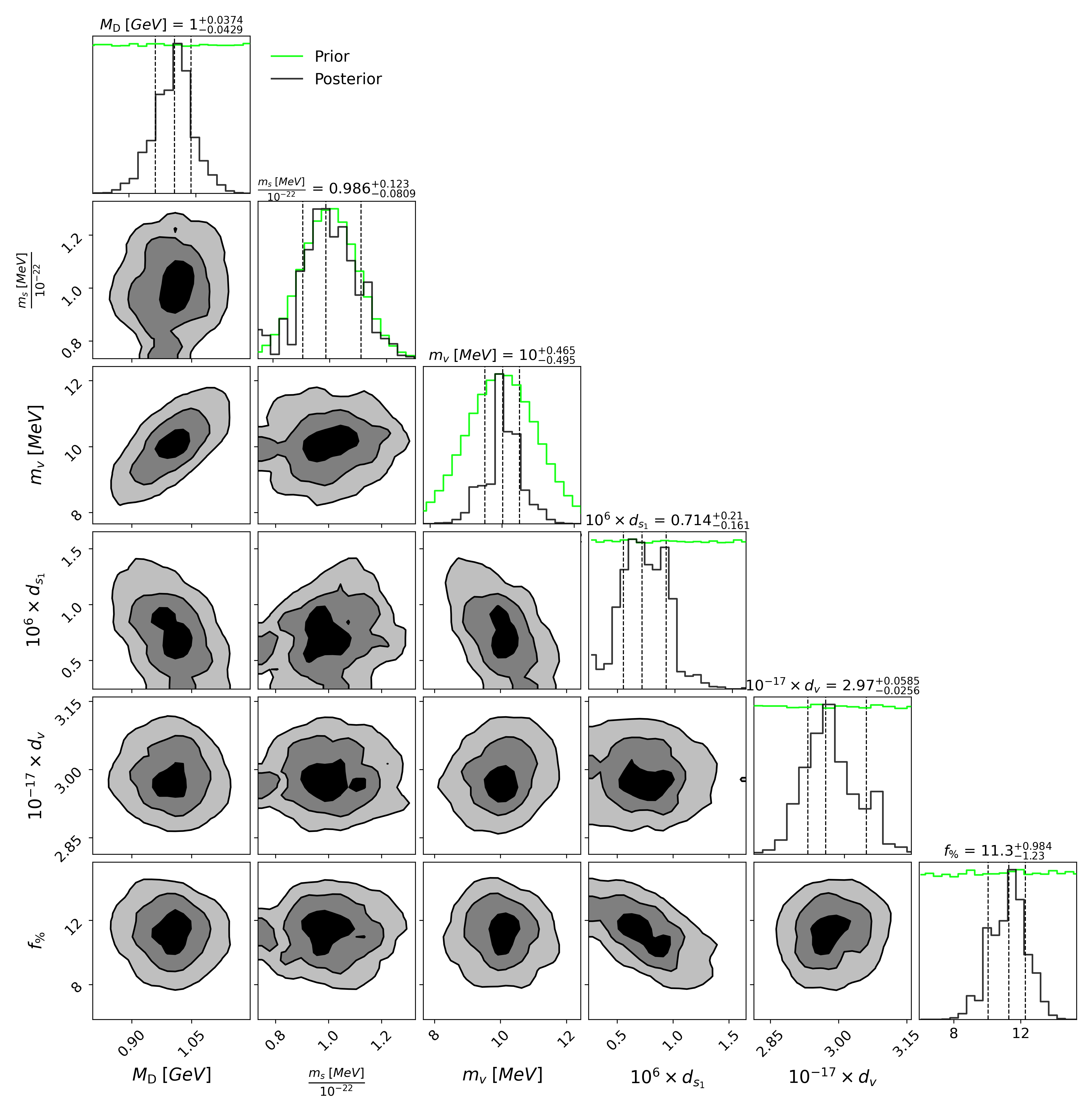}
   \caption{\label{fig:bayesian_linear_MPA1} Same of FIG.~\ref{fig:bayesian_linear_APR4} when DM is combined with BM described by the MPA1 EoS.}
\end{figure*}

\begin{figure*}[h]
\centering
\includegraphics[width=\textwidth]{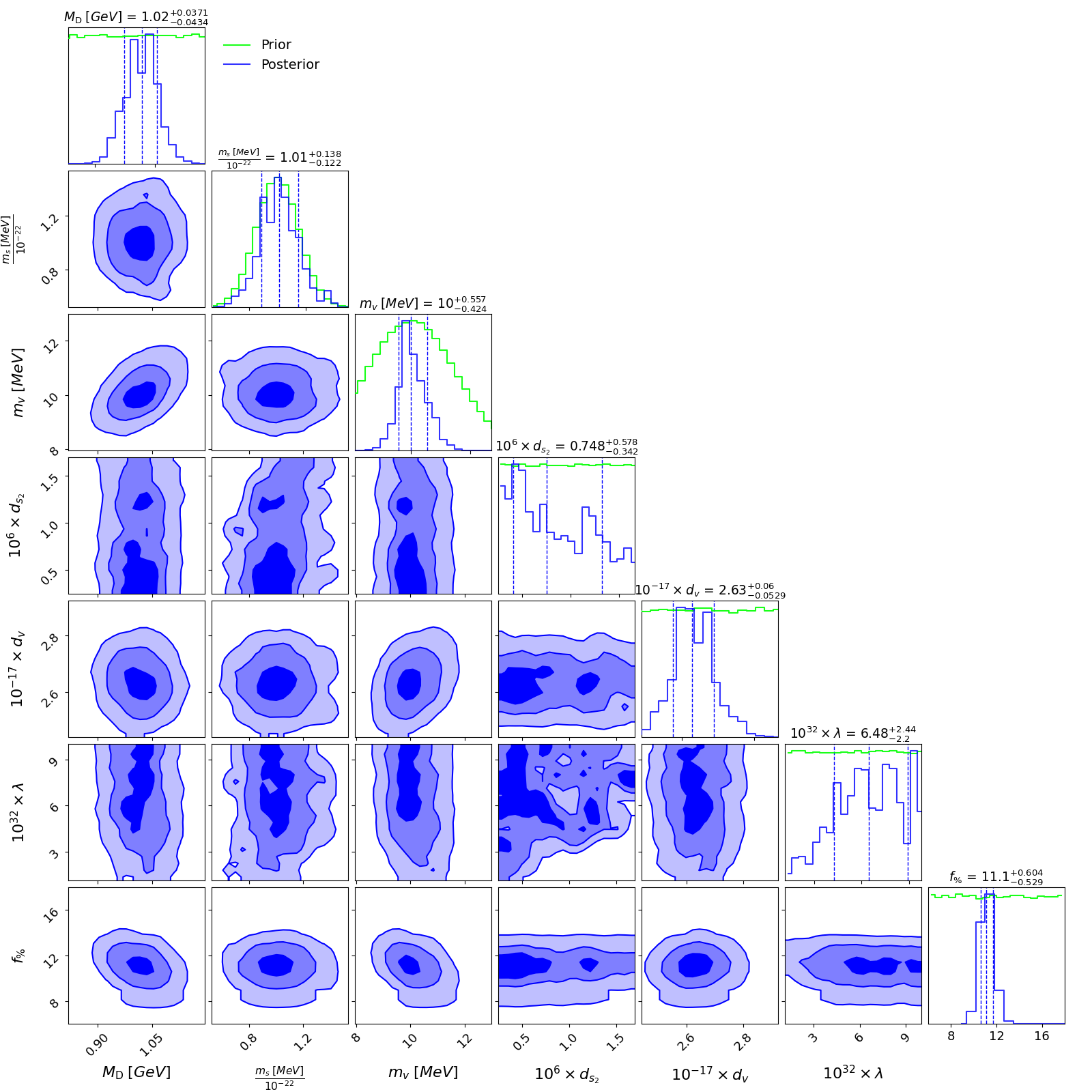}
   \caption{\label{fig:bayesian_quadratic_BSk22} Same of FIG.~\ref{fig:bayesian_quadratic_APR4} when DM is combined with BM described by the BSk22 EoS.}
\end{figure*}

\begin{figure*}[h]
\centering
\includegraphics[width=\textwidth]{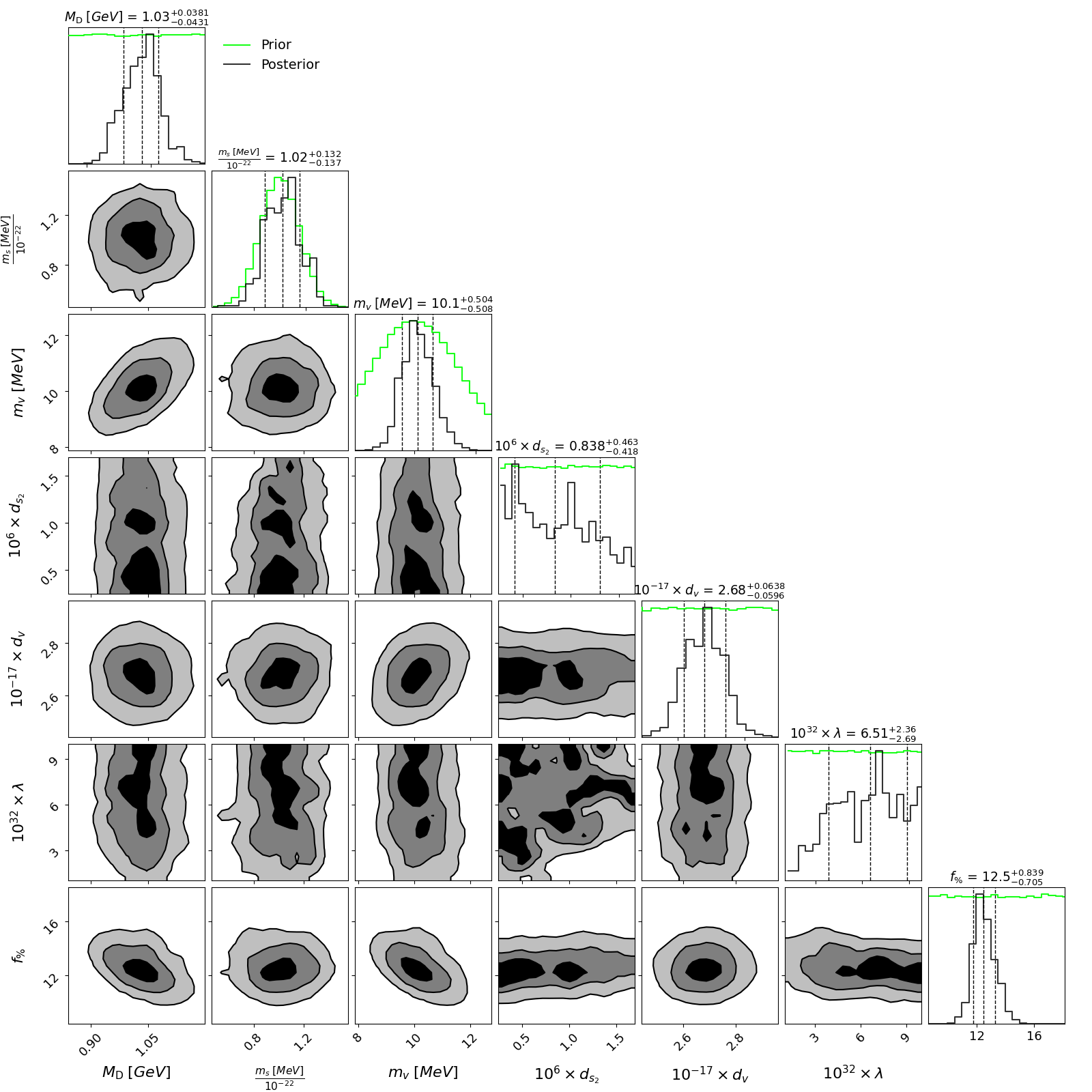}
   \caption{\label{fig:bayesian_quadratic_MPA1} Same of FIG.~\ref{fig:bayesian_quadratic_APR4} when the DM combined with BM described by the MPA1 EoS.}
\end{figure*}

\clearpage

\bibliographystyle{utphys}
\bibliography{bibliography}

\end{document}